\documentclass[a4paper,12pt]{article}
\usepackage[utf8]{inputenc}
\usepackage{cancel}
\usepackage{ulem}
\usepackage{amsfonts}
\usepackage{amssymb}
\usepackage{adjustbox}
\usepackage{graphicx}
\usepackage{amsmath,bm}
\usepackage{enumerate}
\usepackage{comment}
\usepackage{mathtools}
\usepackage{tikz} \usetikzlibrary{calc}
\usepackage[countmax]{subfloat}
\usepackage{xcolor}
\usepackage{float}
\usepackage{color}
\usepackage{siunitx}
\usepackage{array}
\usepackage{booktabs}
\usepackage{tabularx}
\usepackage{here}
\usepackage{cite}
\usepackage{subcaption}

\usepackage{mwe}
\usepackage{multirow}

\usepackage{mathrsfs}
\usepackage{float,epsfig}
\usepackage{dcolumn}
\usepackage{pgfplots}
\usepackage{graphicx}
\usepackage{bm}
\usepackage{amsmath,amssymb,amsthm}
\usepackage[colorlinks=true,linkcolor=blue,citecolor=red]{hyperref}
\usepackage[a4paper,top=2cm,bottom=2cm,left=2cm,right=2cm,marginparwidth=2cm]{geometry}

\definecolor{lime}{HTML}{A6CE39}
\newcommand{\orcidicon}{%
	\begin{tikzpicture}
	\draw[lime, fill=lime] (0,0)
	circle [radius=0.16]
	node[white] {{\fontfamily{qag}\selectfont \tiny ID}};
	\draw[white, fill=white] (-0.0625,0.095)
	circle [radius=0.007];
	\end{tikzpicture}   \hspace{-2mm}
}

\newcommand\orcidHasan{{\href{https://orcid.org/0000-0001-7408-0910}{\orcidicon}}}
\newcommand\orcidKarima{{\href{https://orcid.org/0000-0001-5419-8516}{\orcidicon}}}
\newcommand\orcidZakaria{{\href{https://orcid.org/0009-0006-7872-4713}{\orcidicon}}}

\title{\bf  QPO-Based Bayesian Constraints on Charged Particle Dynamics Around Magnetized Schwarzschild Black Holes
}

\author{
	Z. Ahal\orcidZakaria\!\!$^{1}$\thanks{zakaria.ahal@edu.uiz.ac.ma},  
	H.  El Moumni\orcidHasan\!\!$^1$\thanks{h.elmoumni@uiz.ac.ma }, K. Masmar\orcidKarima\!\!$^{1}$\thanks{k.masmar@uiz.ac.ma (Corresponding author)}\\
{\small $^{1}$ LPTHE, Physics Department, Faculty of Sciences, Ibnou Zohr University, Agadir, Morocco. }
}
\date{\today}
\begin{document} 
	\maketitle
 \begin{abstract}
We study the motion of charged particles with a magnetic dipole moment orbiting a Schwarzschild black hole immersed in an external paraboloidal magnetic field. The interaction between the particle’s intrinsic magnetic moment and the black hole magnetosphere is modeled through a dipole coupling, and the equations of motion are derived using the Hamilton–Jacobi formalism. We analyze equatorial circular orbits, the innermost stable circular orbit, and epicyclic oscillations, showing that the magnetic field strength and coupling parameter produce competing effects on orbital stability and fundamental frequencies. These frequencies are applied to model high-frequency quasi-periodic oscillations within the relativistic precession framework. Using observational QPO data from stellar-mass, intermediate-mass, and supermassive black holes, we perform a Bayesian parameter estimation based on Markov Chain Monte Carlo techniques. The analysis constrains the black hole mass, magnetic field strength, field geometry, coupling parameter, and QPO orbital radius, highlighting the role of magnetospheric interactions in shaping both particle dynamics and timing properties of accreting black holes.
\end{abstract}

	
    \tableofcontents
	
\section{Introduction}
\paragraph{}Recent advances in very long baseline interferometry have culminated in the Event Horizon Telescope (EHT) imaging of the accretion flow around $\textrm{Sgr A}^\star$ and $\textrm{M87}^\star$, providing unprecedented observational support for theoretical predictions and, in particular, for the presence of organized magnetic fields in the immediate vicinity of black holes (BHs) \cite{EventHorizonTelescope:2021bee,EventHorizonTelescope:2024hpu}. Long before this milestone, magnetic fields were already known to play a fundamental role in high-energy astrophysics, as they are detectable and measurable across virtually all classes of compact and non-compact astrophysical objects \cite{silvers2008magnetic,wielebinski1993magnetic,goldreich1969pulsar,blandford1977electromagnetic,beck2016magnetic}. The presence of magnetic fields profoundly influences the dynamics of charged matter, modifies radiation processes, and mediates angular-momentum transport in accretion flows. For such reasons, magnetized environments around black holes have emerged as a privileged arena for probing both high-energy plasma physics and strong-gravity effects. 
Moreover, magnetic fields play a crucial role in explaining many astrophysical processes, such as star and planet formation \cite{mckee2007theory,crutcher2012magnetic}, the acceleration of cosmic rays \cite{blandford2000acceleration,fraschetti2008acceleration,tursunov2020supermassive}, and the launching of jets from matter surrounding black holes \cite{koide2002extraction,marti2015strong,abdujabbarov2014magnetized}. They also contribute to plasma confinement and heating \cite{zaitsev2015particle,biskamp2003magnetohydrodynamic,yamada2010magnetic}. Therefore, astrophysical studies often focus on celestial objects that are potential sources of strong magnetic fields, including magnetars, stars, white dwarfs, pulsars, and the accretion disks of black holes.

Observational studies indicate that magnetic fields are indeed present in the immediate environments of black holes \cite{daly2019black,eatough2013strong}, with strengths that depend on the mass and nature of the accreting system. Estimates span several orders of magnitude: from a few gauss up to $10^8$~G and beyond. For stellar-mass black holes in X-ray binaries, magnetic fields are typically of order $10^8$~G, whereas supermassive black holes (SMBHs) tend to host significantly weaker fields, of order $10^4$~G \cite{eatough2013strong,akiyama2021first,kolovs2023charged}. In particular, $\textrm{Sgr A}^\star$  appears to possess even weaker fields, in the range $10$–$100$~G \cite{piotrovich2010magnetic,Askour:2024nky,Chakhchi:2024tzo}. At such strengths, the magnetic energy density remains subdominant compared to the gravitational binding energy, and therefore does not substantially alter the underlying spacetime geometry. In realistic astrophysical conditions, the exterior geometry of the black hole can thus be well described by the Schwarzschild or Kerr solutions of general relativity.

Analytical magnetic field configurations around black holes have been investigated for several decades. The first exact solution to Maxwell’s equations for a rotating black hole immersed in an asymptotically uniform magnetic field was obtained by Wald \cite{wald1974black}, and later extended to non-rotating geometries \cite{kolovs2023charged,kolovs2019charged,vrba2025charged,kolovs2025charged,kolovs2015quasi}. Alternative configurations have also been proposed, such as the dipole field generated by circular current loops in the Petterson model \cite{petterson1974magnetic}. Other relevant geometries include the split-monopole and paraboloidal configurations introduced in the context of energy extraction mechanisms \cite{blandford1977electromagnetic}. In modern developments, paraboloidal magnetospheres have gained particular interest due to their emergence in general relativistic magnetohydrodynamic (GRMHD) and general-relativistic particle-in-cell (GRPIC) simulations, where they are used to construct realistic models of accretion flows, jet launching regions, and black hole magnetospheres \cite{crinquand2021synthetic,kolovs2020simulations,nakamura2018parabolic,porth2019event,kolovs2023charged}.


Direct measurements of magnetic fields in the vicinity of black holes are not yet accessible. Their presence must therefore be inferred indirectly through their imprint on the dynamics and radiation of matter close to the inner regions of the accretion flow. Magnetic fields accelerate charged particles with intrinsic magnetic moments, modify their orbital motion, and imprint characteristic polarization signatures on the emitted radiation. The recent detection of polarized light from $\textrm{Sgr A}^\star$ by the EHT provides compelling observational support for magnetic field effects operating near the event horizon. In this context, studying the dynamics of magnetized charged particles offers an efficient theoretical framework for probing the electromagnetic environment of black holes. A number of works have examined this problem for different magnetic field configurations and particle models \cite{kolovs2023charged,kolovs2020simulations,vrba2025charged,zhang2025curled}. These studies consistently show that magnetic fields substantially affect the accretion process and the behavior of charged matter in strong gravity. Understanding the interplay between gravity, plasma physics, and electromagnetic fields is therefore essential for constructing realistic models of black hole environments and for interpreting future high-resolution observations.

Furthermore, tidal forces exerted by a black hole on a companion star can trigger Roche lobe overflow and the formation of an accretion disk, as commonly observed in microquasars-binary systems consisting of a stellar-mass black hole and a donor star. Within the disk, angular momentum is transported outward through turbulent and magnetic stresses, while viscous dissipation converts gravitational potential energy into thermal radiation. The innermost regions of the accretion flow are therefore heated to temperatures sufficient to emit copiously in the X-ray band, providing a key observational tracer of matter approaching the event horizon \cite{shahzadi2021epicyclic,jumaniyozov2024circular}. In parallel, part of the accreting plasma can be collimated into relativistic jets, powered by magnetic stresses and rotational energy extraction. Besides, the X-ray flux emitted from accreting black hole systems often exhibits quasi-periodic oscillations (QPOs), which are believed to originate from plasma perturbations and dynamical processes occurring in the strong gravitational field near the compact object \cite{donmez2024perturbing,donmez2024proposing,donmez2024comparison,nishonov2025qpos}. As such, QPOs provide an effective probe of spacetime geometry and of the physical conditions in the inner accretion flow. They have been successfully employed to constrain black hole masses and spins and to test extensions or alternatives to general relativity \cite{remillard2006x,rayimbaev2022radio,abramowicz2001precise,jumaniyozov2025black}. Within this framework, the dynamics of particles endowed with both electric charge and a magnetic dipole moment orbiting a black hole immersed in a parabolic magnetic field constitute a new avenue for exploring the associated observational signatures and potential constraints.


Within such motivations, in this paper, we investigate the dynamics of charged magnetized particles and the associated QPO phenomenology around black holes immersed in a paraboloidal magnetic field. We further perform a Bayesian parameter estimation using Markov Chain Monte Carlo techniques to constrain the black hole mass and magnetospheric parameters within the relativistic precession model, employing observational QPO data from different black hole systems \cite{bouchy2001p,shermatov2025qpos}. 
In addition to estimating the model parameters through the MCMC procedure, we also analyze the correlation matrix of the inferred parameters to investigate possible degeneracies and assess the independence of the resulting constraints.

The paper is organized as follows. In Sec.\ref{section 2}, we introduce the formulation of the external magnetic field surrounding a Schwarzschild black hole and describe the coupling between the particle’s magnetic dipole moment and the external field. In Sec.\ref{section 3}, we analyze the dynamical properties of magnetized charged particles through the effective potential, the stability of circular orbits, and the behavior of the innermost stable circular orbit. We also investigate the radiation properties of the accretion disk and the fundamental oscillation frequencies relevant for QPO modeling within the relativistic precession framework. In Sec.\ref{section 4}, we apply an MCMC analysis to constrain the black hole mass, magnetic field strength, field geometry, coupling parameter, and QPO orbital radius using observational data from stellar-mass, intermediate-mass, and supermassive black holes. Finally, Sec.\ref{conclusion} summarizes our main results and discusses their astrophysical implications.

\section{External Parabolic Magnetosphere of a Schwarzschild Black Hole and Dipole Coupling}\label{section 2}

Herein, we focus on the dynamics of a charged particle endowed with a magnetic dipole moment orbiting a Schwarzschild black hole immersed in an external parabolic magnetic field. The gravitational background is described by the Schwarzschild spacetime, with the line element
\begin{equation}
    ds^2 = -f(r) \,dt^2 + f^{-1}(r) \,dr^2 + r^2 (d\theta^2 + \sin^2{\theta} \,d\phi^2),
\end{equation}
where $f(r) = 1 - 2M/r$. Throughout this analysis we adopt geometrized units and set $M = 1$.

The magnetosphere considered is described by a paraboloidal magnetic field configuration. In this model, the electromagnetic four-potential $A_\mu$ has only a nonvanishing azimuthal component $A_\phi$, consistent with axial symmetry and with the absence of an electric field, such that $A_\mu = (0,\,0,\,0,\,A_\phi)$. 
This heuristic black hole magnetosphere model is motivated by paraboloidal magnetic field solutions that emerge in general relativistic magnetohydrodynamic plasma simulations\cite{crinquand2021synthetic,kolovs2023charged,mckinney2007disc,tchekhovskoy2010black}
\begin{equation}\label{four-vector}
    A_\phi = \frac{1}{2} B_0 r^w (1-| \cos (\theta )| ),
\end{equation}
where, $B_0 \in \mathbb{R}$ represents the magnetic field strength, while the parameter $w \in [0,\,1.25]$ controls the declination of the magnetic field lines. The expression in Eq.\eqref{four-vector} corresponds to a simplified form of the Blandford--Znajek (BZ) paraboloidal model \cite{blandford1977electromagnetic} and does not constitute a vacuum solution of Maxwell's equations in Schwarzschild spacetime. The parameter $w$ interpolates between different configurations: the BZ paraboloidal field at $w=1$ and the split-monopole configuration at $w=0$ \cite{blandford1977electromagnetic,kolovs2023charged}. In GRMHD simulations, a commonly adopted value for describing black hole magnetospheres within jet funnels is $w = 3/4$ \cite{nakamura2018parabolic}, which we shall adopt throughout our investigation.

The absolute value appearing in Eq.\eqref{four-vector} ensures that magnetic field lines diverge across the equatorial plane. In this configuration, the field lines extend below the equatorial plane ($\theta = \pi/2$) while emerging outward above it, yielding a simplified yet physically motivated model of black hole magnetospheres \cite{crinquand2021synthetic,tchekhovskoy2010black,kolovs2023charged}. Despite its heuristic nature, this model retains the most relevant qualitative features of GRMHD magnetospheres, namely the division of the system into three primary regions: the accretion disk, the corona, and the jet funnel. These regions differ primarily in matter density: the accretion disk exhibits the highest density, the jet funnel the lowest, and the corona occupies an intermediate regime characterized by low density and strong turbulent magnetic fields. In contrast, in the jet funnel region the parabolic magnetic field structure dominates over the plasma density. The magnetic field is obtained as follows
\begin{align}
    F_{r \phi} &=  \frac{1}{2} B w \left(1-| \cos (\theta )|\right) r^{w-1},
    & F_{\theta \phi}= \frac{B \sin (\theta ) \cos (\theta ) r^w}{2 | \cos (\theta )|}.
\end{align}

In studying magnetized charged particles, we must also account for the interaction between the external paraboloidal magnetic field and the magnetic dipole moment of the test particle. This interaction is described by the term
\begin{equation}
    2U = -D^{\mu\nu}F_{\mu\nu},
\end{equation}
where $D^{\mu\nu}$ and $F_{\mu\nu}$ denote the dipole polarization tensor and the electromagnetic field tensor, respectively. The polarization tensor is defined as
\begin{equation}
    D^{\alpha\beta} = \eta^{\alpha\beta\sigma\nu} u_\sigma \mu_\nu,
\end{equation}
and satisfies the condition $D^{\mu\nu}u_\nu = 0$, where $u^\mu$ and $\mu^\mu$ correspond to the four-velocity and four-dipole moment of the particle, respectively \cite{jumaniyozov2024collisions,preti2004general}.  At this stage, it is useful to clarify the physical interpretation of the magnetic dipole moment introduced in the model. Indeed, in the present work, the magnetic dipole moment of the test particle
is treated as an effective macroscopic quantity describing the
interaction of a magnetized object with the external magnetosphere.
It should therefore not be interpreted as the quantum spin of a
fundamental particle, but rather as a phenomenological magnetic moment
associated with the internal structure of the orbiting object (e.g.,
magnetized plasma clumps or compact magnetized bodies). Under this assumption, the particle motion is described by modified geodesics
including dipole–field coupling, while spin–curvature effects
described by the Mathisson–Papapetrou–Dixon equations are neglected \cite{Papapetrou:1951pa}.
This approximation is justified in the astrophysical regime considered
here, where electromagnetic interactions dominate over
spin–curvature corrections.
The electromagnetic tensor may be decomposed as
\begin{equation}
    F_{\alpha\beta} = 2u_{[\alpha}E_{\beta]} + \eta_{\alpha\beta\sigma\gamma}u^\sigma B^\gamma,
\end{equation}
where the magnetic field components are given by
\begin{equation}
    B^\alpha = \frac{1}{2}\,\eta^{\alpha\beta\sigma\mu}F_{\beta\sigma}w_\mu,
\end{equation}
with $\eta^{\alpha\beta\sigma\mu} = \sqrt{-g}\,\epsilon^{\alpha\beta\sigma\mu}$, $g = \det(g_{\mu\nu})$ the metric determinant, and $\epsilon^{\alpha\beta\sigma\mu}$ the Levi-Civita symbol. Here $w^\mu$ denotes the four-velocity of a locally non-rotating (proper) observer, expressed as
\begin{equation}
    w^\mu_{\text{proper}} = \left(\frac{1}{\sqrt{-g_{tt}}},0,0,0\right),
\end{equation}
(see~\cite{jumaniyozov2024collisions,murodov2023dynamics}).

In an orthonormal frame, the magnetic field components can be expressed in terms of the electromagnetic tensor as
\begin{equation}
    B^{\hat{i}} = \frac{1}{2}\,\epsilon_{ijk}\,\sqrt{g^{jj}g^{kk}}\,F_{jk}.
\end{equation}

 Next, for simplicity and to ensure a stable configuration, we assume that the magnetic dipole moment of the charged particle is aligned with the external parabolic magnetic field lines of the black hole. Under this assumption, the dipole moment is effectively frozen into the local magnetic field, and the four-dipole moment of the particle takes the form
\begin{equation}
    \mu^{\hat{i}} = (0,\,\mu^{\hat{\theta}},\,0),
\end{equation}
which yields a stable equilibrium configuration for magnetized particles in the external field~\cite{jumaniyozov2024collisions,rayimbaev2024particles}. The interaction term therefore becomes
\begin{equation}
    D^{\mu\nu}F_{\mu\nu} = 2\,\mu^{\hat{\alpha}}B_{\hat{\alpha}}.
\end{equation}

For the paraboloidal magnetic field model, the nonvanishing orthonormal components read
\begin{align}
    B^{\hat{r}} = \frac{1}{2}B_0\,| \cos (\theta )|\,\sec{\theta}\, r^{w-2} \quad \text{ and }\quad
    B^{\hat{\theta}} = \frac{1}{2}B_0\,w\,\sqrt{f(r)}\,\csc{\theta}\, r^{w-2}(|\cos{\theta}|-1).
\end{align}
Thus,
\begin{equation}
    D^{\mu\nu}F_{\mu\nu} = \mu B_0\, w\,\sqrt{f(r)}\,\csc{\theta}\, r^{w-2}(|\cos{\theta}|-1).
\end{equation}

We may finally express the interaction potential as
\begin{equation}\label{coupling}
    U(r,\theta) = -\beta\,w\,\sqrt{f(r)}\,\left(|\cos (\theta )|-1\right)\csc{\theta}\, r^{w-2},
\end{equation}
where $\beta = \mu B_0 / 2$ is the magnetic coupling parameter.

In astrophysical scenarios such as a neutron star orbiting a supermassive black hole, one has $\mu = (1/2)B_{\rm NS}R_{\rm NS}^3$, where $B_{\rm NS}$ and $R_{\rm NS}$ denote the neutron star surface magnetic field and radius, respectively. The coupling parameter then reads~\cite{narzilloev2021dynamics}
\begin{equation}
    \beta \simeq \frac{11}{250}
    \left(\frac{B_{\rm NS}}{10^{12}\,\mathrm{G}}\right)
    \left(\frac{R_{\rm NS}}{10^{6}\,\mathrm{cm}}\right)^3
    \left(\frac{B_{\rm ext}}{10\,\mathrm{G}}\right)
    \left(\frac{m_{\rm NS}}{M_\odot}\right)^{-1},
\end{equation}
where $m_{\rm NS}$ is the neutron star mass. The paraboloidal magnetic field, together with the coupling parameter $\beta$ provides a suitable framework to investigate the dynamics and stability of magnetized particles within accretion disks surrounding black holes.

Having established the external magnetic field configuration and the dipole coupling, we now turn to the dynamics of a magnetized charged particle in such an environment. The interplay between gravity, magnetic fields, and dipole interactions affects the orbital structure close to the black hole, and we explore its implications through the corresponding equations of motion.

\section{Magnetized charged particles Dynamics in a Parabolic Black Hole Magnetosphere} \label{section 3}

\subsection{Effective potential and equation of motion}

Using the Hamilton-Jacobi formalism, we derive the equations of motion for a magnetized charged particle in the external parabolic magnetic field of the Schwarzschild black hole, including the dipole-field interaction. The Hamilton-Jacobi equation takes the form \cite{preti2004general}
\begin{equation}
    g^{\mu\nu}\left(\frac{\partial S}{\partial x^\mu} - q A_\mu\right)\left(\frac{\partial S}{\partial x^\nu} - q A_\nu\right)
    = -\left(m + U\right)^2,
\end{equation}
where $m$ denotes the rest mass of the particle, which we set to $m=1$ hereafter. The action $S$ describes the motion of the particle, and we restrict our analysis to circular orbits in the equatorial plane, $\theta=\pi/2$, yielding
\begin{equation}
    S = -Et + L\phi + S_r.
\end{equation}
This allows for the separation of variables and reduces the radial equation of motion to
\begin{equation}\label{dotr}
    \dot{r}^2 = E'^2 - V_{\rm eff}(r;L'),
\end{equation}
where $E' = E/m$ and $L' = L/m$ are the specific energy and angular momentum, and the effective potential takes the form
\begin{equation}
    V_{\rm eff}(r;L') = f(r)\left[\left(1 + U(r)\right)^2 + \left(\frac{L' - B\, r^w(1-|\cos\theta|)}{r\sin\theta}\right)^2\right].
\end{equation}

The parameter $L'$ corresponds to the specific angular momentum of the particle, where positive (negative) values describe counterclockwise (clockwise) motion as viewed from above the black hole. The quantity $B = qB_0/2m$ denotes the magnetic parameter associated with the external field, and its sign determines the orientation of the magnetic field along the $z$-axis: $B>0$ corresponds to field lines emerging above the equatorial plane, whereas $B<0$ corresponds to the opposite orientation. The interaction term $U(r)$ contributes an additional asymmetry to the potential, breaking the symmetry of $V_{\rm eff}$ with respect to $L'$. Further discussion of the symmetric case $U=0$ can be found in \cite{kolovs2023charged}. Thus, the sign of $\beta$ is located in Eq.~\eqref{coupling}, indicating whether the two magnetic field lines are aligned (anti-aligned) where $\beta>0$ ($\beta<0$).

Alternatively, the equations of motion may be derived from the Lagrangian~\cite{jumaniyozov2024collisions}
\begin{equation}
    \mathcal{L} = \frac{1}{2}\left(m+U(r)\right)g_{\mu\nu}u^\mu u^\nu - \frac{1}{2}kU(r) + qA_\mu u^\mu,
\end{equation}
from which the specific constants of motion follow:
\begin{equation}
    -E' = \left(m+U(r)\right)g_{tt}\dot{t} + qA_t,
\end{equation}
\begin{equation}\label{Lphi}
    L' = \left(m+U(r)\right)g_{\phi\phi}\dot{\phi} + qA_\phi.
\end{equation}

\begin{figure}[!ht]
    \centering
    \includegraphics[width=0.5\textwidth]{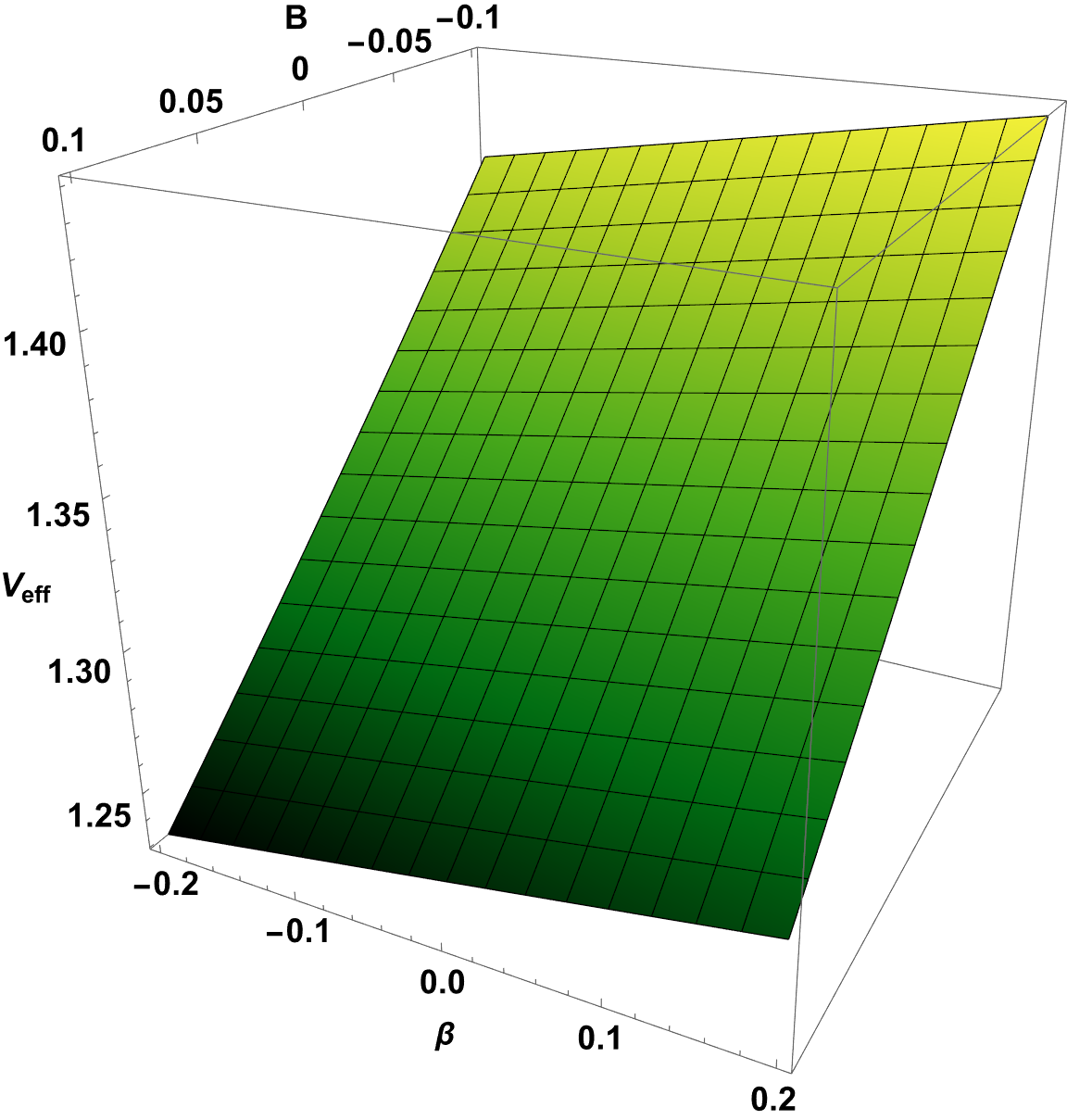}
    \caption{\footnotesize{\it Effective potential as a function of the external magnetic field strength $B$ and the coupling parameter $\beta$, for fixed values $r=6$ and $L'=6$. The parameters $B$ and $\beta$ exhibit opposite qualitative effects: increasing $B$ lowers the potential, whereas increasing $\beta$ raises it.}}
    \label{effective_potential}
\end{figure}

Fig.\ref{effective_potential} displays the response of the effective potential to variations in the external magnetic field strength $B$ and in the dipole coupling parameter $\beta$. 

For fixed orbital radius and angular momentum in the equatorial plane, the two parameters influence the potential in opposite directions: an increase in $B$ lowers $V_{\rm eff}$ and therefore favors tighter orbits, whereas an increase in $\beta$ raises $V_{\rm eff}$, effectively acting against magnetic confinement. The decrease induced by $B$ is considerably stronger than the increase due to $\beta$, indicating that the external field dominates over the dipole–field interaction term in this configuration.

\newpage

\subsection{Circular orbits and the innermost stable circular orbit}

\paragraph{} The circular motion of charged magnetized particles around a magnetized Schwarzschild black hole is governed by the structure of the effective potential $V_{\rm eff}$. Circular orbits are defined by the absence of radial motion and acceleration, i.e.,
\begin{equation}
\dot r = \ddot r = 0 \,,
\end{equation}
which is equivalent to the condition
\begin{equation}
\partial_r V_{\rm eff} = 0 .
\end{equation}
The stability of circular orbits is determined by the second radial derivative of the effective potential:
\begin{equation}
\partial_{r}^2 V_{\rm eff} > 0 \quad (\text{stable}), \qquad
\partial_{r}^2 V_{\rm eff} < 0 \quad (\text{unstable}),
\end{equation}
while the marginally stable configuration, corresponding to the innermost stable circular orbit (ISCO), is given by
\begin{equation}
\partial_r V_{\rm eff} = \partial_{r}^2 V_{\rm eff} = 0 .
\end{equation}

Under these conditions, the specific energy of a charged magnetized particle follows directly from Eq.\eqref{dotr} and can be written as
\begin{equation}
E'^2 = V_{\rm eff}(r; L', B', \bm{\kappa}, \beta) ,
\label{eff,energy}
\end{equation}
which defines the boundaries of allowed particle motion. Depending on the magnetic field strength and the inclination of the orbit, these boundaries may be closed or open along the $z$-axis. Since our analysis focuses on the accretion disk and corona regions—where matter density dominates over magnetic turbulence—the boundary structure is sensitive to both the external magnetic field and the dipole coupling parameter. Similar boundary behavior has been discussed for uniform magnetic fields in Ref.~\cite{stuchlik2019magnetized}. In particular, particle escape may occur either from the corona toward the jet funnel \cite{kolovs2020simulations} or from the inner edge of the accretion disk \cite{ripperda2022black}. In contrast, within the jet funnel itself, the boundaries are typically open, allowing particles to propagate outward with relativistic velocities.

The specific angular momentum associated with circular orbits is obtained from the condition $\partial_r V_{\rm eff}=0$ and takes the form
\begin{equation}
L'_{\pm} =
-\frac{B r^w \left[r f'(r)+(w-2) f(r)\right]
\pm r^3 \sqrt{B^2 w^2 f(r)^2 r^{2 w-6}+X(r)}}
{2 f(r)-r f'(r)} ,
\end{equation}
where
\begin{equation}
X(r) =
\frac{(1+U(r)) \left[2 f(r)-r f'(r)\right]
\left[(1+U(r)) f'(r)+2 f(r) U'(r)\right]}{r^3} .
\end{equation}
Here, primes denote derivatives with respect to the radial coordinate $r$. The $+$ and $-$ branches correspond to stable and unstable circular orbits, respectively.

\begin{figure}[!ht]
    \centering
    \includegraphics[width=0.5\linewidth]{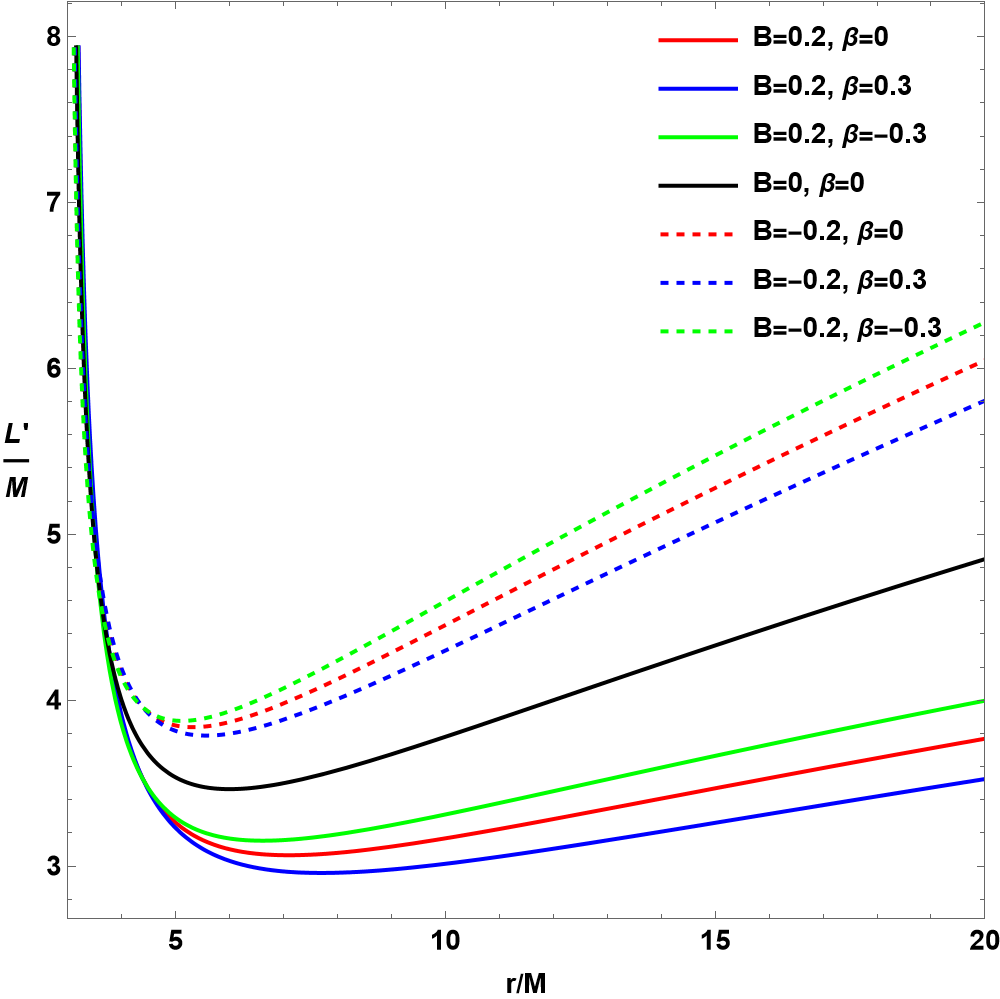}
    \caption{\footnotesize{\it Radial dependence of the specific angular momentum $L'$ for circular orbits of charged magnetized particles for different values of the magnetic field strength $B$ and coupling parameter $\beta$.}}
    \label{fig:angular fct of radii}
\end{figure}

Figure~\ref{fig:angular fct of radii} displays the radial profiles of the critical specific angular momentum for stable circular orbits. Dashed curves correspond to negative values of the magnetic field strength, while thick curves represent cases with $B \geq 0$. The solid black curve denotes the unmagnetized case ($B=\beta=0$). A decrease in the coupling parameter $\beta$ leads to an increase in $L'$, and this trend persists for both magnetic polarities, indicating that the dipole interaction provides a systematic correction to the Lorentz-force-dominated dynamics.

\begin{figure}[!ht]
    \centering
    \includegraphics[width=0.5\linewidth]{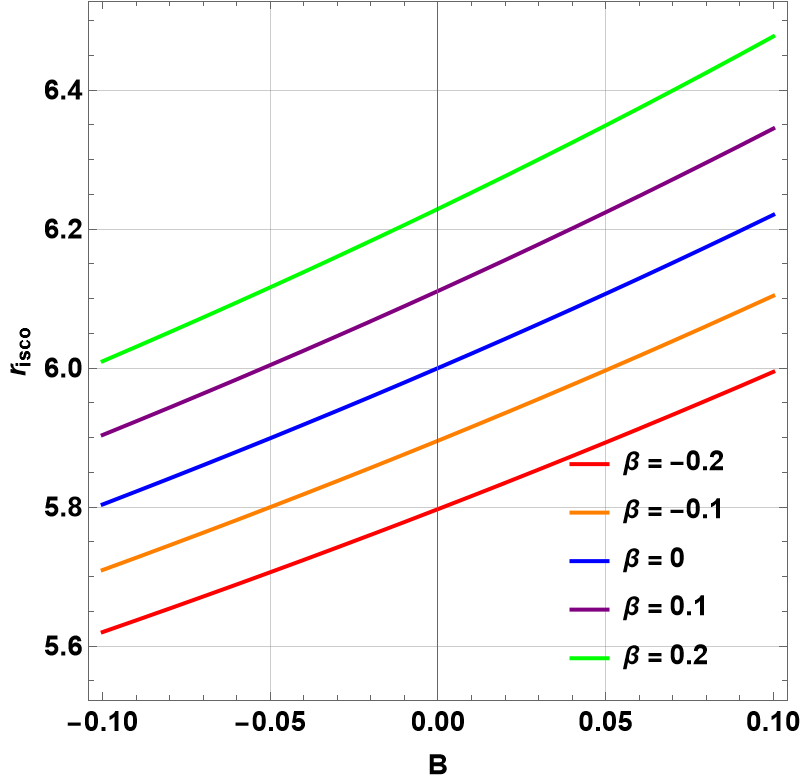}
    \caption{ISCO radius $r_{\rm ISCO}$ as a function of the magnetic field strength $B$ for different values of the coupling parameter $\beta$.}
    \label{fig:isco graph}
\end{figure}

Further, the ISCO radius is determined by the inflection point condition $\partial_{r}^2V_{\rm eff}=0$. Its dependence on the magnetic field strength and coupling parameter is shown in Fig.\ref{fig:isco graph}.

As illustrated in Fig.~\ref{fig:isco graph}, both the external magnetic field and the dipole coupling affect the ISCO in a similar manner. Positive values of $(B,\beta)$ shift the ISCO outward, increasing the inner radius of the accretion disk, whereas negative values draw the ISCO closer to the event horizon.

\begin{figure}[!ht]
    \centering
    \includegraphics[width=\linewidth]{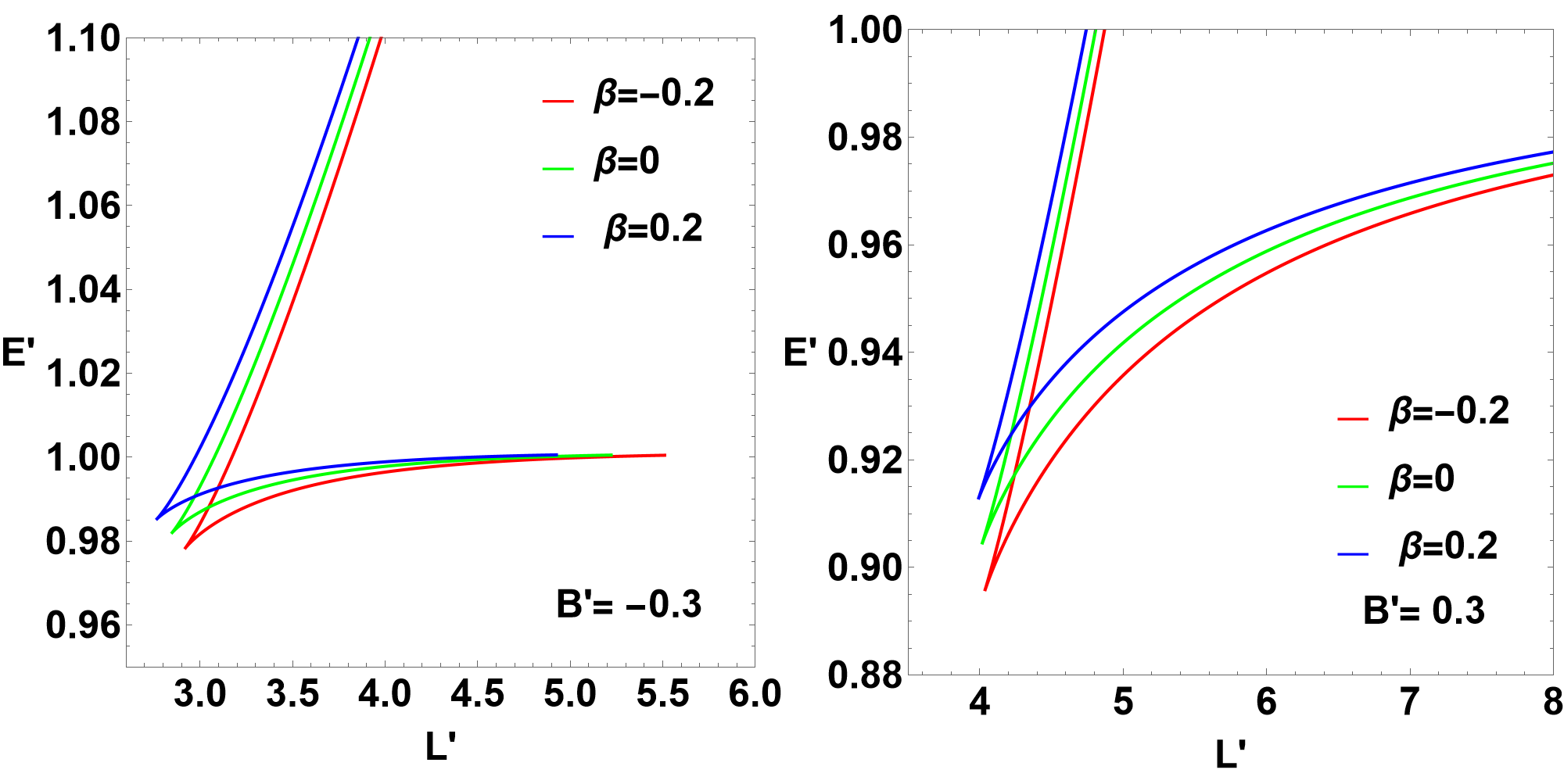}
    \caption{\footnotesize{\it Specific energy $E'$ as a function of the specific angular momentum $L'$ for different values of the coupling parameter $\beta$ and magnetic field strength $B$.}}
    \label{fig:energy fct of angular}
\end{figure}

At the ISCO, the particle possesses well-defined values of specific energy and angular momentum. Fig.\ref{fig:energy fct of angular} shows the dependence of the specific energy on the angular momentum for different values of $\beta$ and magnetic field polarity.

Both $E'$ and $L'$ vary with the coupling parameter. As $\beta$ decreases, the specific energy slightly decreases while the angular momentum increases. For moderate magnetic field strengths (e.g. $B=0.3$), these variations remain subdominant.

\begin{figure}[!ht]
    \centering
    \subfloat[]{\includegraphics[width=\textwidth]{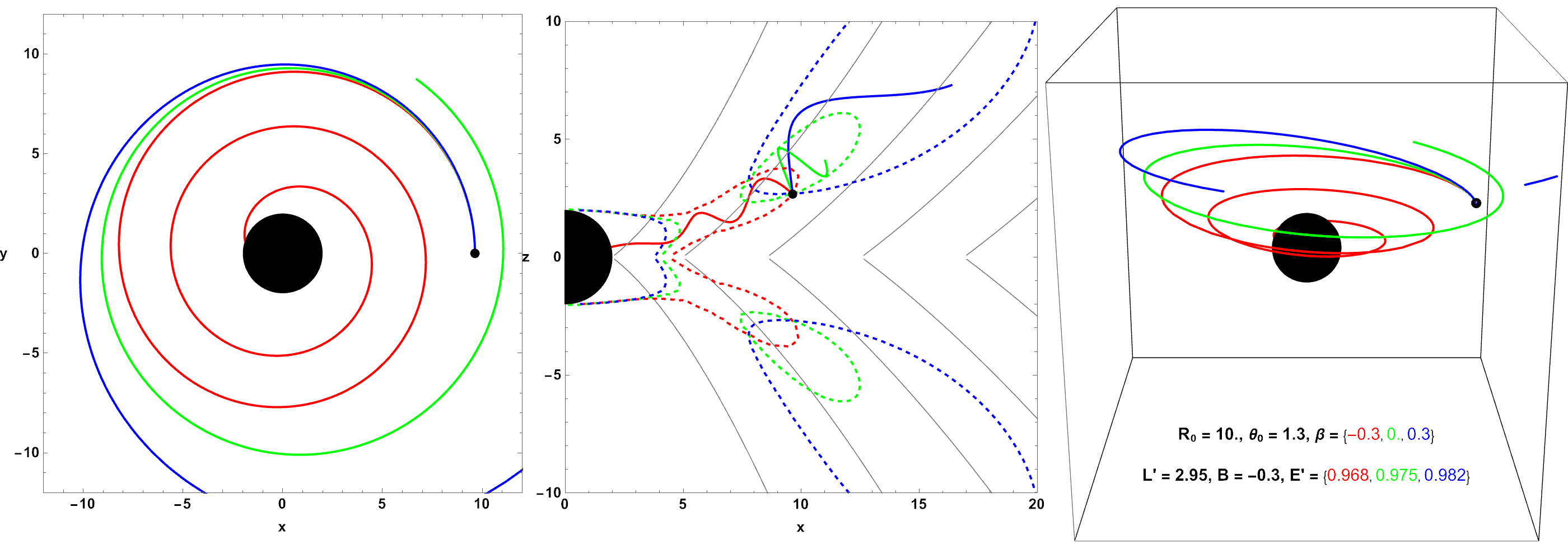}}\\
    \subfloat[]{\includegraphics[width=\textwidth]{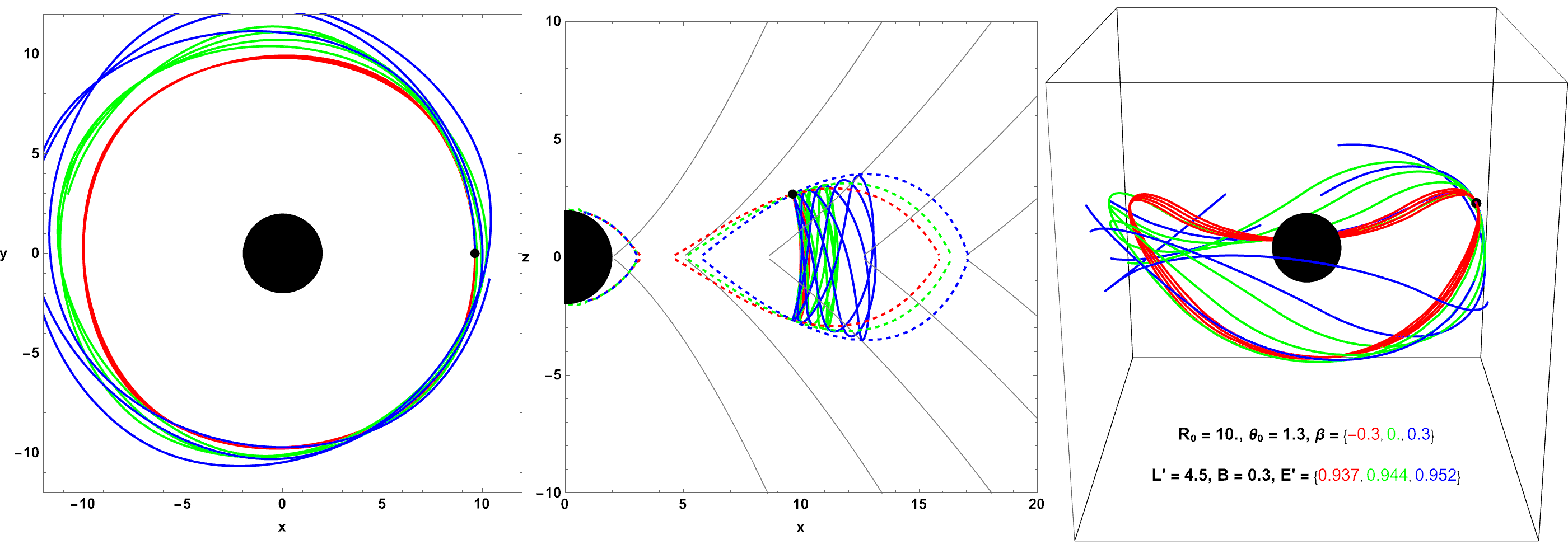}}\\
    \subfloat[]{\includegraphics[width=\textwidth]{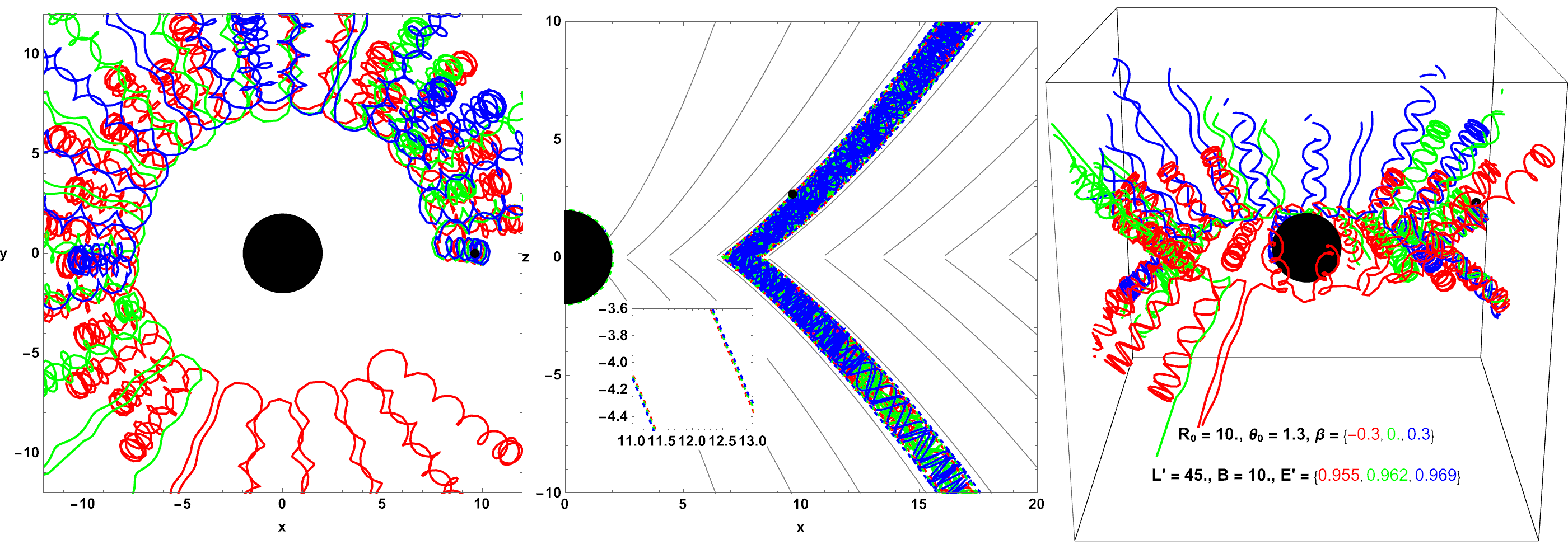}}
    \caption{\footnotesize{\it Trajectories of charged magnetized particles around a magnetized black hole for different magnetic field strengths and coupling parameters. The black disk represents the black hole, while gray curves denote magnetic field lines.}}
    \label{weak MF}
\end{figure}

As shown in Ref.\cite{kolovs2023charged}, a negative (positive) magnetic field polarity corresponds to an attractive (repulsive) Lorentz force. The inclusion of a magnetic dipole moment significantly modifies this picture. Representative particle trajectories for weak and strong magnetic fields are displayed in Fig.\ref{weak MF} for different values of $\beta$.

In the weak magnetic field regime, particle motion remains largely regular. For an attractive Lorentz force and negative $\beta$, particles are confined near the black hole within a single connected boundary. As $\beta$ increases, this boundary splits into two distinct regions, allowing long-lived orbital motion without capture or escape. In the repulsive Lorentz-force regime, increasing $\beta$ shifts the boundaries outward, enlarging the allowed region of motion.

In contrast, strong magnetic fields induce chaotic behavior, characterized by oscillatory motion along the $z$-axis guided by magnetic field lines. In this regime, the allowed region forms a tunnel-like structure aligned with the magnetic field, and increasing $\beta$ further enhances the particle energy and the width of the accessible region.

Overall, both the external magnetic field and the dipole coupling parameter play a crucial role in shaping particle dynamics, orbital stability, and the location of the ISCO.

 The modification of circular orbits and the shift of the ISCO induced by the external magnetic field and the dipole coupling have direct observational consequences. Small perturbations of stable circular orbits give rise to radial and vertical epicyclic oscillations, whose characteristic frequencies are naturally associated with quasi-periodic oscillations (QPOs) observed in the X-ray flux of accreting black hole systems. Since both the Lorentz force and the dipole–field interaction alter the effective potential and the stability properties of circular motion, they are expected to leave distinct imprints on the epicyclic frequencies and their ratios.

\subsection{Radiation properties}
In order to investigate the radiative properties of an accretion disk composed of magnetized charged particles orbiting a black hole immersed in an external parabolic magnetic field, we employ the Novikov--Thorne thin-disk model. This framework assumes a geometrically thin ($h \ll r$) and optically thick accretion disk in hydrodynamic equilibrium, where radiative cooling efficiently balances viscous heating.

Within this model, the radiation emitted by the disk is primarily generated near its inner edge, close to the ISCO, where relativistic effects and magnetic interactions are strongest. The presence of charged particles carrying a magnetic dipole moment and interacting with the black hole magnetosphere modifies the disk energetics and, consequently, its observable radiation signatures. 
The radiative flux of the accretion disk is given by the Novikov--Thorne prescription
\begin{equation}\label{eq flus radiation}
\mathcal{F}(r)=\frac{-\dot{M}}{4\pi\sqrt{g}}
\frac{\Omega_{,r}}{(E'-\Omega L')^2}
\int_{r_{\rm isco}}^{r}(E'-\Omega L')L'_{,r}\,dr,
\end{equation}
where $\dot{M}$ denotes the mass accretion rate, which is set to unity for simplicity. The quantity $\sqrt{g}=\sqrt{-g_{tt}g_{rr}g_{\phi\phi}}$ is the determinant of the induced metric in the equatorial plane. The specific energy $E'$ and angular momentum $L'$ correspond to the prograde branch of circular orbits.
 The Keplerian angular velocity reads
\begin{equation}
\Omega = \frac{\dot{\phi}}{-g^{tt}E'},
\end{equation}
where $\dot{\phi}=d\phi/d\tau$ follows from Eq.~\eqref{Lphi}, and the factor $(-g^{tt}E')^{-1}$ accounts for gravitational redshift effects.

Assuming blackbody emission, the temperature profile of the disk is
\begin{equation}
T(r)=\left(\frac{\mathcal{F}(r)}{\sigma}\right)^{1/4},
\end{equation}
where $\sigma$ is the Stefan--Boltzmann constant. 
\begin{figure}[!ht]
    \centering
    \subfloat[]{\includegraphics[width=0.48\linewidth]{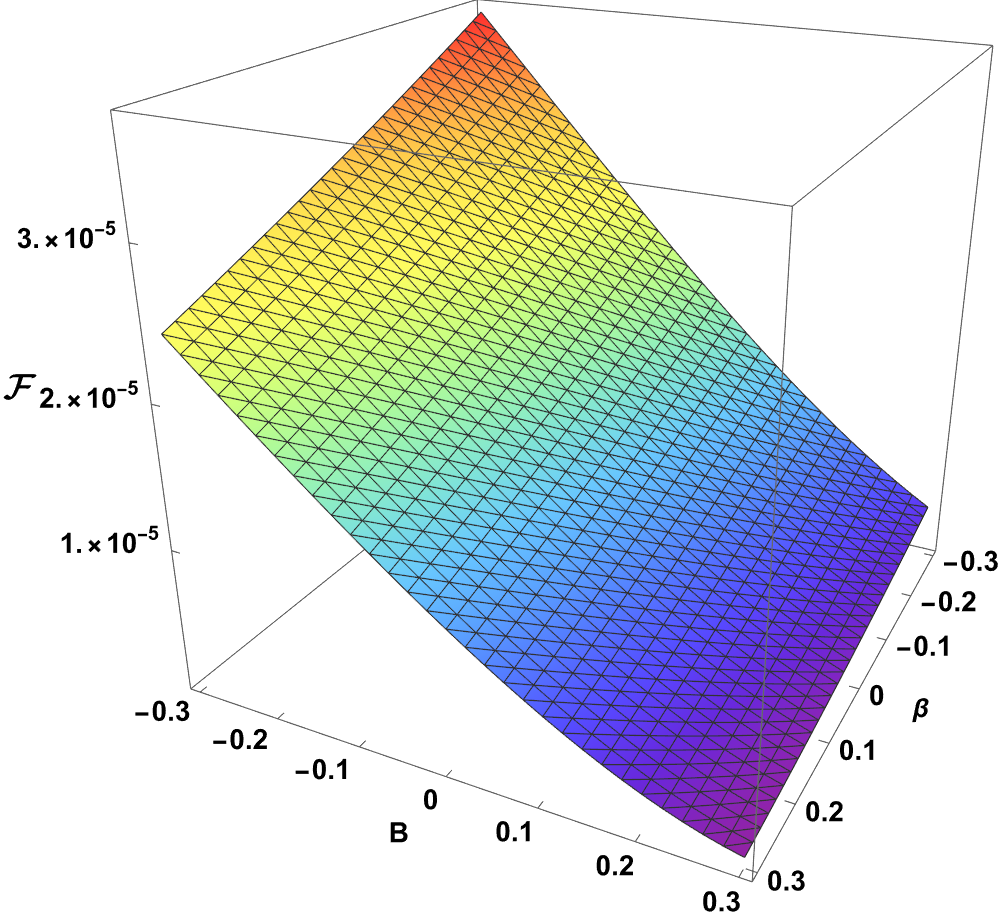}}
    \subfloat[]{\includegraphics[width=0.46\linewidth]{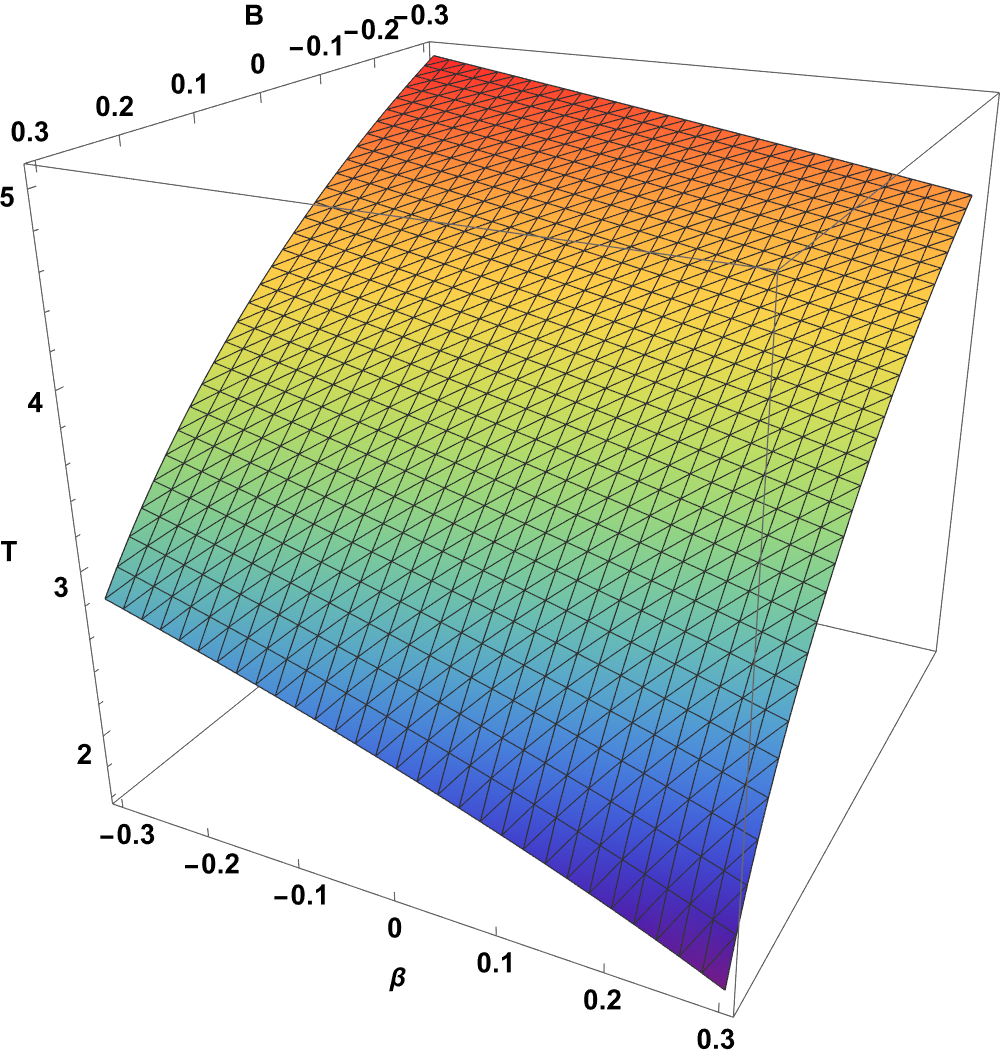}}
    \caption{\footnotesize{\it Radiating flux of accretion disk  within panel (a)  and Temperature of the disk in panel (b) against $B'$ and $\beta$ in a fixed $r=10$ }.}
    \label{fig:flux}
\end{figure}

Fig.\ref{fig:flux} displays the dependence of the radiative flux and temperature on the magnetic field strength $B$ and the coupling parameter $\beta$ at a fixed radius $r=10$. Negative values of $B$ enhance both the flux and temperature, while increasing $\beta$ suppresses them. These trends reflect the competing roles of the Lorentz force and dipole--field coupling in redistributing orbital energy within the disk.

\begin{figure}[!ht]
    \centering
     \subfloat[$B=-0.3\,, \beta=-0.3$]{\includegraphics[width=0.33\linewidth]{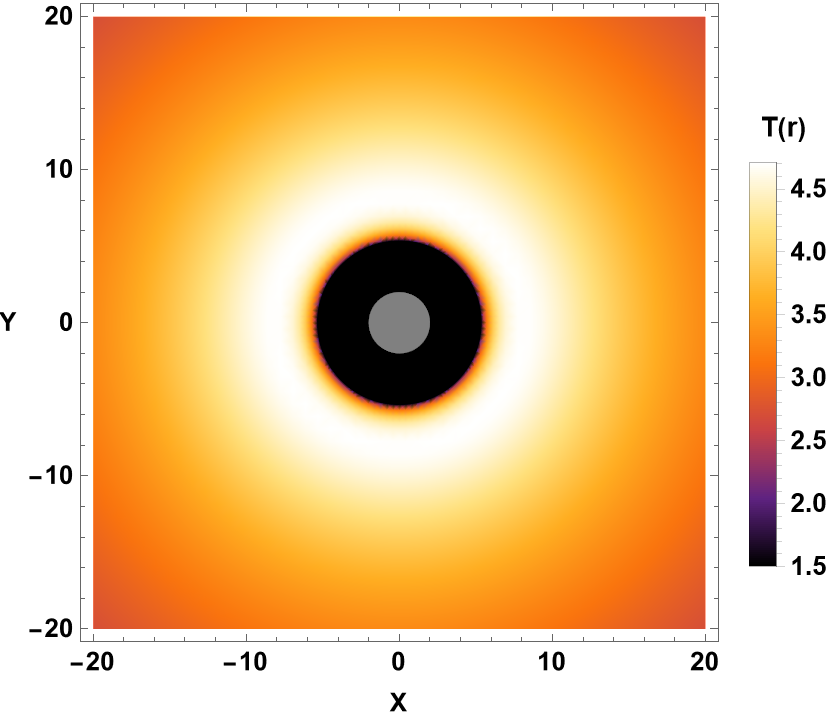}}
    \subfloat[$B=0.3\, ,\beta=-0.3$]{\includegraphics[width=0.33\linewidth]{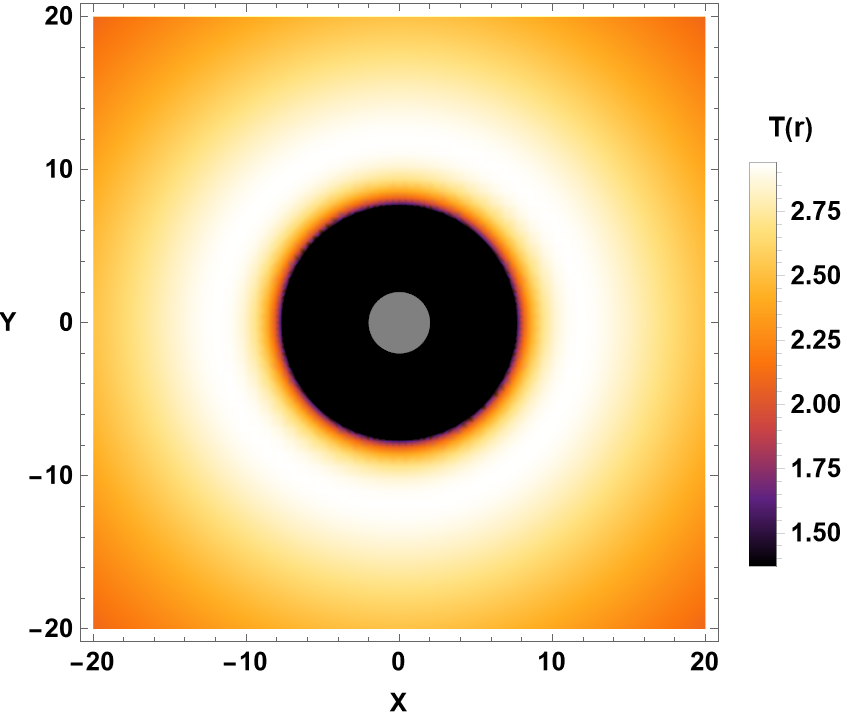}}
     \subfloat[$B=0.3\, ,\beta = 0.3$]{\includegraphics[width=0.33\linewidth]{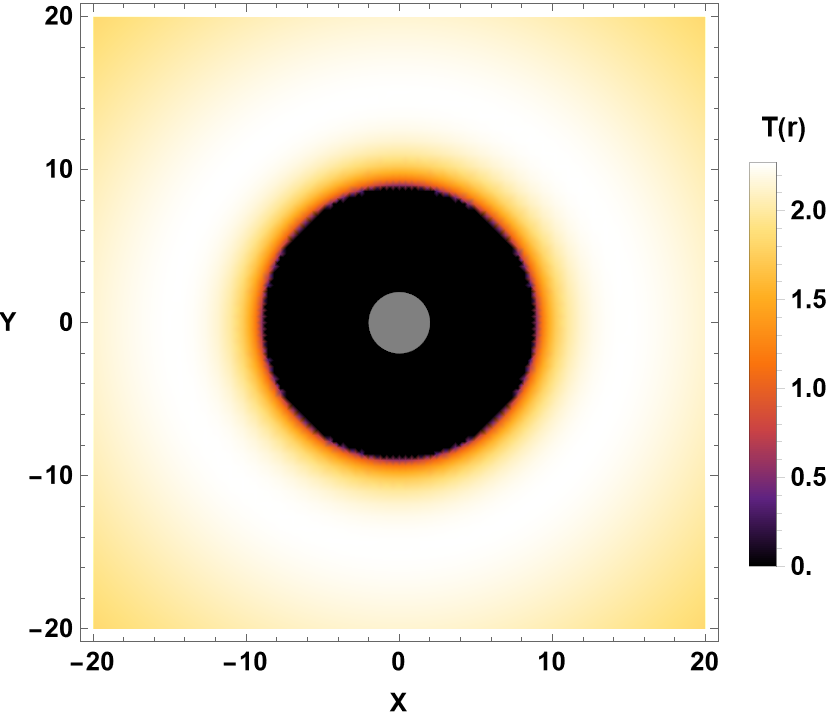}}\\
    \subfloat[$B= -0.3\, ,\beta=0.3$]{\includegraphics[width=0.33\linewidth]{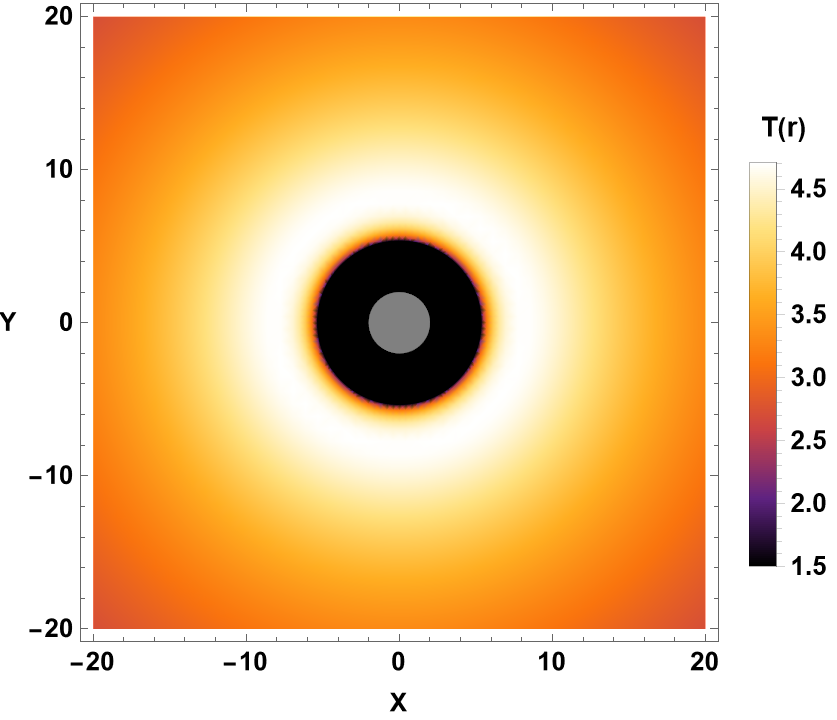}}
    \subfloat[$B=0\,,\beta=0$]{\includegraphics[width=0.33\linewidth]{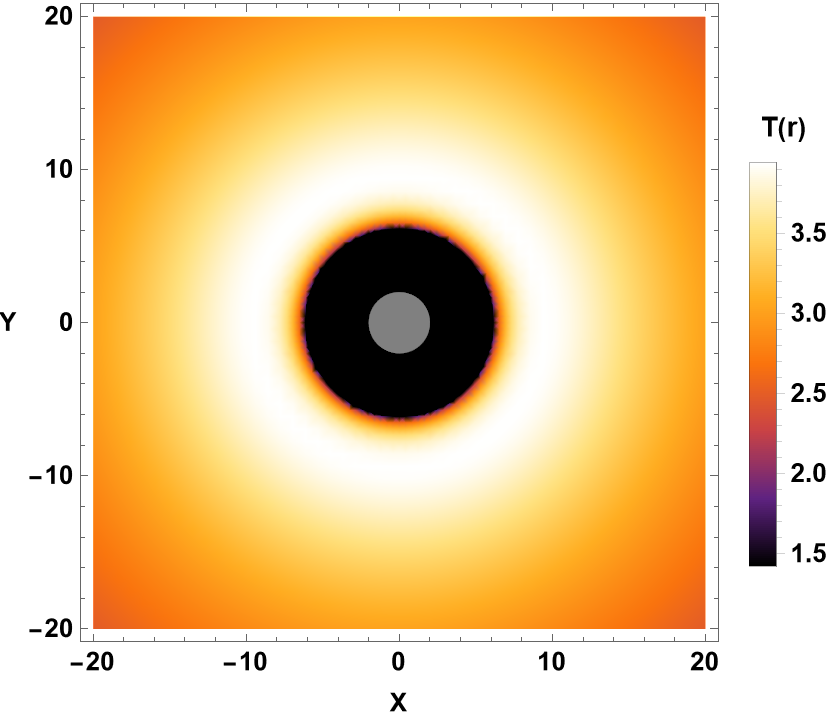}}
    \caption{\footnotesize{\it Density plot of the temperature profile of the accretion disk around the black hole for different parabolic MF strengths $B$ and coupling parameters $\beta$.}}
    \label{fig:density plot}
\end{figure}

The two-dimensional temperature distribution of the disk is illustrated in Fig.\ref{fig:density plot}. 

Negative magnetic polarity concentrates high-temperature regions closer to the ISCO, while positive values of $B$ and $\beta$ shift both the ISCO radius and the thermal peak outward. This behavior is consistent with the dynamical analysis presented in the previous section.

The differential luminosity observed at infinity is given by
\begin{equation}
\frac{d\mathcal{L}_\infty}{d\ln r} = 4\pi r\sqrt{g}\,E'\,\mathcal{F}(r),
\end{equation}
which directly reflects the radial distribution of emitted radiation.

Assuming blackbody radiation, the spectral luminosity measured at infinity reads
\begin{equation}
\nu\mathcal{L}_{\nu,\infty} = \frac{60}{\pi^3}
\int_{r_{\rm isco}}^{\infty}
\frac{\sqrt{g}E'}{M^2}
\frac{(u^t y)^4}
{\exp\!\left[\frac{u^t y}{(M^2\mathcal{F})^{1/4}}\right]-1}
\,dr,
\end{equation}
where $y=h\nu/(kT_\star)$ and $T_\star$ is defined through the Stefan--Boltzmann law $\sigma , T_\star = \frac{M_0}{4 \pi M^2},$.

\begin{figure}[h!]
    \centering
    \includegraphics[width=0.5\linewidth]{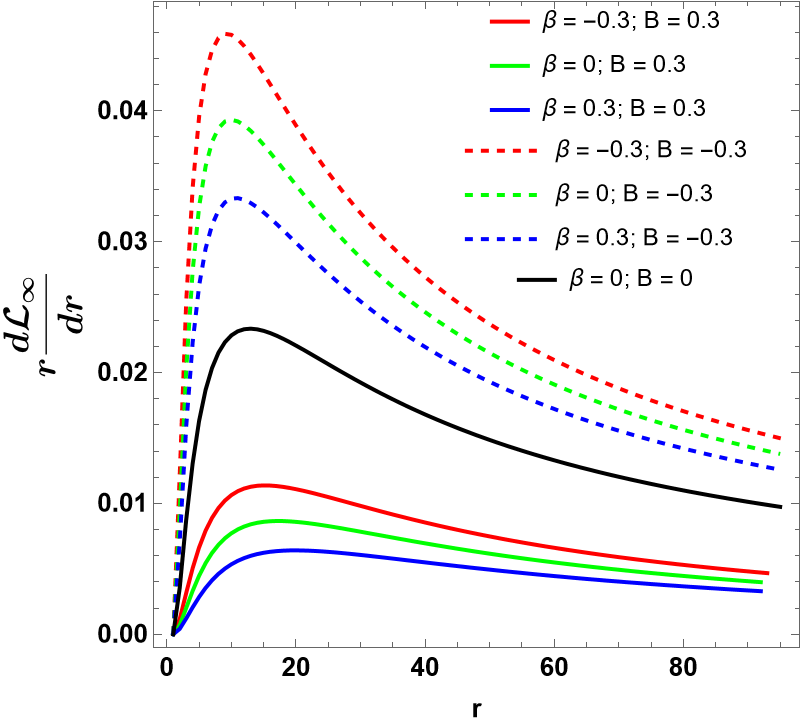}
    \caption{\footnotesize{\it Differential luminosity of accretion disk within different strength of $B$ and $\beta$.}}
    \label{fig:variation of Luminosity}
\end{figure}

Moreover, Fig.\ref{fig:variation of Luminosity} illustrates the radial dependence of the differential luminosity of the accretion disk for different values of the magnetic field strength $B$ and the coupling parameter $\beta$.

  The luminosity exhibits behavior consistent with that observed for the radiative flux and temperature profiles. In particular, for negative values of $B$, the differential luminosity is significantly enhanced in the vicinity of the black hole and rapidly decreases with increasing radius. Conversely, for positive values of $B$, the luminosity remains comparatively lower and shows only a mild radial decay.

The coupling parameter $\beta$ further modulates this behavior by altering the effective energy budget of the magnetized particles. Increasing $\beta$ leads to a suppression of the emitted luminosity, reflecting the additional energy stored in the dipole--field interaction. These results demonstrate that both the polarity of the external magnetic field and the strength of the dipole coupling play a crucial role in shaping the radiative efficiency and observational appearance of magnetized accretion disks.

 In the following section, we therefore investigate the epicyclic dynamics of magnetized charged particles in a parabolic black hole magnetosphere and analyze the resulting QPO frequencies.

\subsection{Fundamental frequencies}

The dynamics of particles orbiting a black hole is strongly governed by the properties of circular orbits, in particular those located near the (ISCO), which defines the inner edge of the accretion disk. At this radius, particles follow circular motion corresponding to an extremum of the effective potential in the equatorial plane ($\theta=\pi/2$).  When a particle is subjected to a small perturbation away from this equilibrium configuration, it undergoes oscillatory motion around the circular orbit. These oscillations are characterized by three fundamental frequencies: the Keplerian (orbital) frequency, the radial epicyclic frequency, and the vertical epicyclic frequency. These frequencies play a central role in modeling quasi-periodic oscillations (QPOs) observed in accreting black hole systems.  

In this section, we derive the fundamental frequencies governing the motion of charged particles endowed with a magnetic dipole moment orbiting a Schwarzschild black hole immersed in a paraboloidal magnetic field.
\subsection{Keplerian frequency}

For circular motion in the equatorial plane, the Keplerian frequency describes the azimuthal orbital motion of a charged magnetized particle at a given radius. This frequency is directly linked to the dynamics of the accretion disk and therefore to its radiative efficiency. Observationally, the Keplerian frequency is commonly associated with high-frequency QPOs detected in X-ray binaries.

The Keplerian frequency, as measured by a distant observer at infinity, is defined as
\begin{equation}
    \Omega_\phi \equiv \Omega_K = \frac{\omega_\phi}{-g^{tt} E'} ,
\end{equation}
where the factor $(-g^{tt}E')^{-1}$ accounts for the gravitational redshift. The local angular velocity $\omega_\phi$ follows from Eq.\eqref{Lphi} and is given by
\begin{equation}
    \omega_\phi = g^{\phi\phi} \left( \frac{L' - B r^w \left(1-|\cos\theta|\right)}{1+U(r,\theta)} \right).
\end{equation}

\begin{figure}[!ht]
    \centering
    \includegraphics[width=0.5\linewidth]{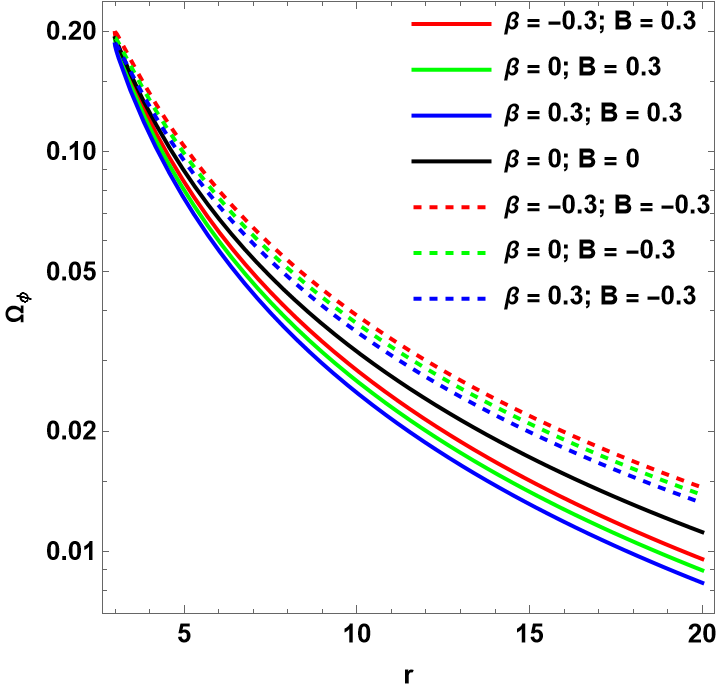}
    \caption{\footnotesize{\it Radial profiles of Keplerian frequencies for a charged particle with dipole moment orbiting a magnetized BH within different strengths of $B$ and $\beta$.}}
    \label{fig:keplerian frequen}
\end{figure}

In the remainder of this analysis, we restrict our attention to equatorial motion ($\theta=\pi/2$). The radial profiles of the Keplerian frequency for different values of the magnetic field strength $B$ and coupling parameter $\beta$ are displayed in Fig.\ref{fig:keplerian frequen}.
The thick curves correspond to non-negative magnetic field strengths ($B \geq 0$), while dashed curves represent negative values ($B < 0$). The Keplerian frequency decreases with increasing magnetic field strength. The coupling parameter $\beta$ produces a qualitatively similar effect, although its influence is generally weaker. As a result, the combined effects of $(+B,+\beta)$ and $(-B,-\beta)$ lead to lower and higher Keplerian frequencies, respectively.
\subsection{Harmonic oscillations}

Small perturbations around circular orbits lead to harmonic oscillations in both the radial and vertical directions. These perturbations can be written as
\begin{equation}
r = r_0 + \delta r, \qquad \theta = \frac{\pi}{2} + \delta\theta ,
\end{equation}
where $r_0$ denotes the radius of the circular orbit.

\begin{figure}[!ht]
    \centering
    \includegraphics[width=0.5\linewidth]{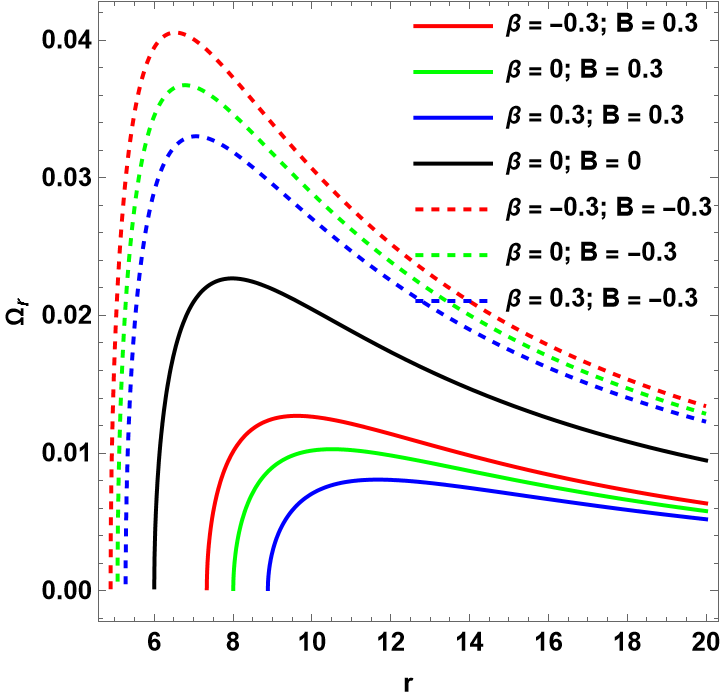}
    \caption{\footnotesize{\it Radial dependence of the radial frequency for a charged particle with dipole moment orbiting a magnetized BH within different strengths of $B$ and $\beta$.}}
    \label{fig:radial frequen}
\end{figure}

The evolution of these perturbations is governed by
\begin{equation}
    \ddot{\delta i} + \omega_i^2 \delta i = 0,
\end{equation}
with $i=\{r,\theta\}$ and overdots denoting derivatives with respect to the particle’s proper time. The radial and vertical epicyclic frequencies are given by
\begin{align}
    \omega_r^2 = \frac{\partial^2 V_{\rm eff}}{\partial r^2}, \quad
    \omega_\theta^2 = \frac{1}{r^2 f(r)} \frac{\partial^2 V_{\rm eff}}{\partial \theta^2}.
\end{align}
The corresponding frequencies measured by a distant observer are
\begin{equation}
    \Omega_{r,\theta} = \frac{\omega_{r,\theta}}{-g^{tt} E'}.
\end{equation}
In Fig.\ref{fig:radial frequen}, we depict the radial dependence of the radial epicyclic frequency $\Omega_r$ for different values of $B$ and $\beta$.

This figure reveals that for non-negative magnetic field strengths $(B \ge 0)$, the radial epicyclic frequency $\Omega_r$ is reduced, whereas it is enhanced for $B<0$. Likewise, negative values of the coupling parameter $\beta$ increase $\Omega_r$, while positive values lead to its suppression.

The shift in the starting point of $\Omega_r$ reflects variations in the ISCO radius. Increasing $B$ and $\beta$ moves the ISCO outward, while negative values bring it closer to the black hole horizon.

\subsection{QPO studies in the relativistic precession model}

In this subsection, we investigate the characteristic frequencies of twin-peak quasi-periodic oscillations (QPOs) produced by charged magnetized particles orbiting a magnetized black hole within the framework of the relativistic precession (RP) model. The RP model is a well-established approach that attributes QPOs to relativistic corrections in particle orbits in the strong gravitational field near a black hole. Originally introduced by Stella and Vietri, the model was developed to explain kilohertz twin-peak QPOs observed in neutron star systems within the frequency range $0.2$–$1.25$~kHz \cite{stella1999correlations,nishonov2025qpos}. Unlike resonance-based models \cite{torok2005orbital,abramowicz2004orbital,smith2021confrontation,tremaine2014dynamics,ahal2025modeling}, the RP model does not require exceptionally strong magnetic fields or preferred resonance radii. Instead, it describes QPOs as arising from small perturbations of plasma blobs moving along slightly eccentric and tilted trajectories within the accretion disk.

The fundamental frequencies governing particle motion can be translated into the corresponding frequencies measured by distant observers located at spatial infinity. These observed frequencies are given by
\begin{equation}
 \nu_i = \frac{c^3}{2\pi G M} \frac{\omega_i}{-g^{tt} E'}, \qquad i = \{r,\theta,\phi\},
\end{equation}
where $G$ and $c$ denote the gravitational constant and the speed of light, respectively, and $M$ is the black hole mass.

Within this framework, QPOs originate from small perturbations of particle trajectories governed by the intrinsic frequencies of the spacetime. In the RP model, the upper high-frequency QPO is identified with the Keplerian frequency of particles orbiting in the innermost regions of the accretion disk,
 $
\nu_U = \nu_\phi,
$ while the lower high-frequency QPO is associated with the relativistic periastron precession of slightly eccentric orbits, defined as
 $
\nu_L = \nu_\phi - \nu_r,
$ 
where $\nu_r$ is the radial epicyclic frequency. Since $\nu_r < \nu_\phi$, this difference captures the relativistic precession of particle orbits \cite{ahal2025modeling}. In addition, horizontal branch oscillations (HBOs) are commonly attributed to nodal precession, described by the frequency difference $\nu_\phi - \nu_\theta$, which reflects vertical oscillations induced by frame dragging in rotating black hole spacetimes. However, in the present study, the black hole is non-rotating, and frame dragging is therefore absent. In purely geodesic motion, this implies $\nu_\theta = \nu_\phi$. Nevertheless, the presence of an external magnetic field modifies the particle dynamics and breaks this degeneracy, leading to $\nu_\theta \neq \nu_\phi$ even in the Schwarzschild background \cite{ahal2025modeling}. As a consequence, the periastron precession frequency $\nu_L = \nu_\phi - \nu_r$ becomes the dominant contributor to the lower high-frequency QPO over a wide range of magnetic field strengths and coupling parameters \cite{stella1999correlations,hazarika2025signatures,ahal2025modeling}.

\begin{figure}[!ht]
    \centering
    \includegraphics[width=0.5\linewidth]{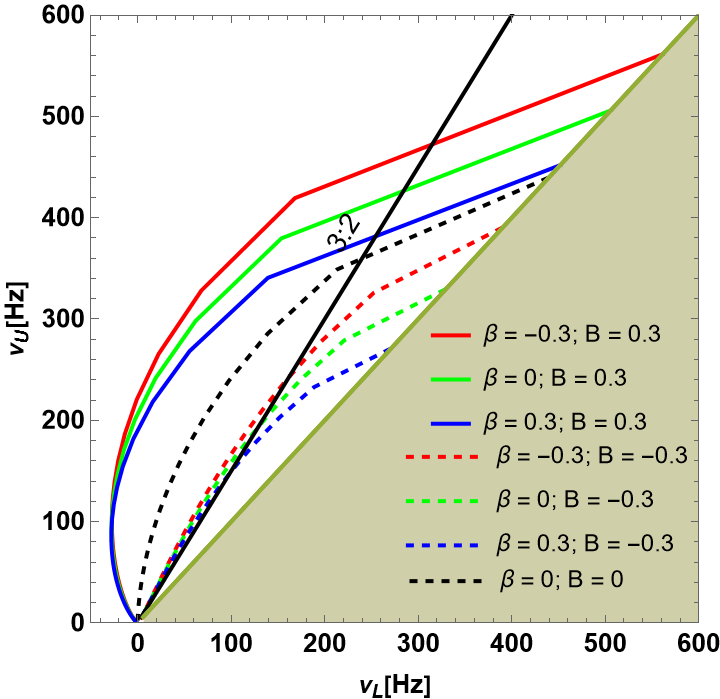}
    \caption{\footnotesize{\it Upper and lower QPO frequency correlation predicted by the relativistic precession model for different values of the magnetic field strength $B$ and coupling parameter $\beta$.}}
    \label{fig:upper and lower}
\end{figure}

In several microquasar systems, high-frequency QPOs observed in X-ray binaries often appear in the Fourier power spectrum with a characteristic $3\!:\!2$ frequency ratio. This feature is commonly interpreted as a signature of resonance phenomena occurring in the inner regions of the accretion disk \cite{Abramowicz:2001bi,torok2005orbital,Motta:2013wwa}. In Fig.\ref{fig:upper and lower}, we present the correlation between the upper and lower QPO frequencies generated in the vicinity of a Schwarzschild black hole immersed in a paraboloidal magnetic field. In constructing this diagram, we assume a black hole mass of $M = 5\,M_\odot$. The twin-peak QPOs are modeled within the RP framework.

This figure shows that the characteristic frequencies shift systematically under the combined influence of the black hole magnetic field strength and the dipole–field coupling parameter, while preserving the typical $3\!:\!2$ frequency ratio. In particular, a positive (negative) value of $B$ leads to an increase (decrease) in both QPO frequencies, whereas the coupling parameter $\beta$ exhibits the opposite behavior: increasing (decreasing) $\beta$ results in lower (higher) frequencies. This demonstrates that magnetic interactions provide an efficient mechanism for modulating QPO frequencies without destroying their characteristic ratio.

Theframework developed above establishes a direct connection between the black hole magnetosphere, particle dynamics, and the resulting QPO frequencies. To quantitatively confront these predictions with observational data and constrain the relevant physical parameters, we now turn to a statistical analysis based on the Markov Chain Monte Carlo (MCMC) method.

\section{Bayesian MCMC Constraints from QPO Observations}\label{section 4}

In this section, we constrain the physical parameters of charged particles endowed with magnetic dipole moments orbiting a Schwarzschild black hole immersed in an external parabolic magnetic field. To this end, we employ observational high-frequency quasi-periodic oscillation (HF QPO) data from X-ray binaries spanning three distinct black hole mass scales: stellar-mass, intermediate-mass, and supermassive black holes.

For the stellar-mass category, we consider the microquasars GRO~J1655--40, GRS~1915+105, and XTE~J1550--564 \cite{remillard2002evidence,morgan1997rxte,strohmayer2001discovery},
 but, for simplicity, the present analysis is performed in the Schwarzschild spacetime. This choice allows us to isolate the impact of magnetic interactions and dipole-field coupling on the dynamics of charged particles without introducing additional effects associated with
black hole rotation. In a Kerr spacetime, frame dragging would modify
the orbital motion and the associated epicyclic frequencies entering
the relativistic precession (RP) model through Lense–Thirring precession \cite{borah2025black}.

We note that several of the microquasars considered in this work, including GRS~1915+105 and GRO~J1655-40, are widely believed to host rapidly rotating black holes. However, the primary goal of the present study is to investigate how magnetospheric interactions influence
particle dynamics and the resulting QPO phenomenology. Adopting a
non-rotating background, therefore, provides a minimal baseline framework in which the magnetic Lorentz force and dipole coupling can be clearly identified.

The QPO signals from these systems were detected using high-time-resolution X-ray instruments capable of resolving variability down to microsecond scales \cite{gendreau2016neutron}. These oscillations appear as narrow peaks in the power density spectrum, extracted via Fourier analysis of the photon count rate \cite{van1989fourier,nishonov2025qpos}. As a representative intermediate-mass black hole, we include the ultraluminous X-ray source M82~X--1, whose HF QPOs indicate a black hole mass in the range $10^2$--$10^3\,M_\odot$ \cite{fiorito2004m82,stuchlik2015mass}. Finally, we also analyze the supermassive black hole $\textrm{Sgr A}^\star$, where QPOs are observed in the millihertz regime due to its large mass and low accretion luminosity \cite{nishonov2025qpos}.

The observational QPO frequencies and corresponding black hole masses for all considered sources are summarized in Tab.\ref{QPOs}.

\begin{table}[!ht]
\centering
\caption{\footnotesize{\it Observed QPOs frequencies and masses for selected sources.}}
\label{QPOs}
\adjustbox{max width=\textwidth}{
\begin{tabular}{lccccc}
\toprule
\vspace{1mm}
Source & $\nu_U$ [Hz] & $\Delta \nu_U$ [Hz] & $\nu_L$ [Hz] & $\Delta \nu_L$ [Hz] & Mass [$M_\odot$] \\
\hline
\vspace{2mm}
GRS 1915+105 & 168   & $\pm 3$           & 113   & $\pm 5$            & $12.4^{+2.0}_{-1.8}$ \\
\vspace{2mm}
GRO J1655-40 & 451   & $\pm 5$           & 298   & $\pm 4$            & $5.4 \pm 0.3$ \\
\vspace{2mm}
XTE J1550-564  & 276   & $\pm 3$           & 184   & $\pm 5$            & $12.4^{+2.0}_{-1.8}$ \\
\vspace{2mm}
M82 X-1      & 5.07  & $\pm 0.06$        & 3.32  & $\pm 0.06$         & $415 \pm 63$ \\
\vspace{2mm}
$\textrm{Sgr A}^\star$       & 1.445 & $\pm 0.16\ \text{mHz}$ & 0.886 & $\pm 0.04\ \text{mHz}$ & $(4.1 \pm 0.6)\times 10^{6}$ \\
\bottomrule
\end{tabular}
}
\end{table}
To determine the best-fit values of the black hole mass, QPO orbital radius, magnetic field strength, dipole coupling parameter, and magnetic field geometry parameter, we employ a Bayesian inference framework based on Markov Chain Monte Carlo (MCMC) techniques. The posterior probability distribution of the model parameters is given by
\begin{equation}
\mathcal{P}(\theta|\mathcal{D},\mathcal{M}) =
\frac{P(\mathcal{D}|\theta,\mathcal{M})\,\pi(\theta|\mathcal{M})}
{P(\mathcal{D}|\mathcal{M})},
\end{equation}
where $\theta=\{M,\mathcal{B},r/M,\beta,w\}$ denotes the set of free parameters, $\pi(\theta)$ represents the prior distribution, and $P(\mathcal{D}|\theta,\mathcal{M})$ is the likelihood function.

We adopt Gaussian priors for all parameters within physically motivated bounds (see Table~\ref{priors}),
\begin{equation}
\pi(\theta_i)\propto
\exp\!\left[-\frac{1}{2}\left(\frac{\theta_i-\theta_{0,i}}{\sigma_i}\right)^2\right],
\qquad
\theta_{\mathrm{low},i}<\theta_i<\theta_{\mathrm{high},i}.
\end{equation}

\begin{table}[!ht]
\centering
\caption{\footnotesize{\it The Gaussian priors ($\mu$ is the mean value and $\sigma$ the variance) of the BH parameters from QPOs of the sources.}}\label{priors}
\adjustbox{max width=\textwidth}{
\begin{tabular}{lcccccccccc}
\toprule
 & \multicolumn{2}{c}{XTE J1550-564} & \multicolumn{2}{c}{GRO J1655-40} & \multicolumn{2}{c}{GRS 1915+105} & \multicolumn{2}{c}{Sgr \( A^* \)} & \multicolumn{2}{c}{M82 X-1} \\
\cmidrule(lr){2-3} \cmidrule(lr){4-5} \cmidrule(lr){6-7} \cmidrule(lr){8-9} \cmidrule(lr){10-11}
Parameter & \(\mu\) & \(\sigma\) & \(\mu\) & \(\sigma\) & \(\mu\) & \(\sigma\) & \(\mu\) & \(\sigma\) & \(\mu\) & \(\sigma\) \\
\midrule
\(M/M_{\odot}\) & 12.47 & 0.06 & 5.4 & 0.2 & 10.8 & 0.1 & \(4.01 \times 10^6\) & \(0.30 \times 10^6\) & 438.13 & 63.69 \\
\( \mathcal{B}\)   & 3.1 & 0.4 & 2.5 & 0.3 & 2.5 & 0.3 & 1.12 & 0.3 & 3.4 & 0.2 \\
\(r/M\)   & 4.4 & 0.40 & 5.6 & 0.5 & 6.77 & 0.4 & \(4.9\) & 0.4 & 6.1 & 0.2 \\
\(\mathcal{\beta}\)   & \(-3.1\) & 0.4 & \(-1.1\) & 0.3 & \(1.6\) & 0.4 & \(-1.2\) & 0.3 & \(-3.1\) & 0.4 \\
\(w\)   & 0.32 & 0.04 & 0.01 & 0.002 & 0.07 & 0.02 & \(0.9\) & 0.05 & 0.13 & 0.05 \\
\bottomrule
\end{tabular}
}
\end{table}

The likelihood is constructed from the observed upper and lower HF QPO frequencies following the relativistic precession (RP) model. The total log-likelihood reads
\begin{equation}
\log\mathcal{L}=\log\mathcal{L}_U+\log\mathcal{L}_L,
\end{equation}
with
\begin{align}
\log\mathcal{L}_U &= -\frac{1}{2}\sum_i
\frac{\left(\nu^{i}_{U,\mathrm{obs}}-\nu^{i}_{\phi,\mathrm{th}}\right)^2}
{\sigma^{i\,2}_{U,\mathrm{obs}}},\\
\log\mathcal{L}_L &= -\frac{1}{2}\sum_i
\frac{\left(\nu^{i}_{L,\mathrm{obs}}-\nu^{i}_{L,\mathrm{th}}\right)^2}
{\sigma^{i\,2}_{L,\mathrm{obs}}}.
\end{align}
Here, $\nu_U=\nu_\phi$ is the Keplerian frequency and
$\nu_L=\nu_\phi-\nu_r$ denotes the periastron precession frequency. We perform the MCMC sampling using the affine-invariant ensemble sampler \texttt{emcee} \cite{foreman2013emcee}. For each source, we generate chains of approximately $10^5$ samples to ensure convergence and robust exploration of the parameter space.

\begin{figure}[H]
    \centering
    \subfloat[]{\includegraphics[width=0.45\linewidth]{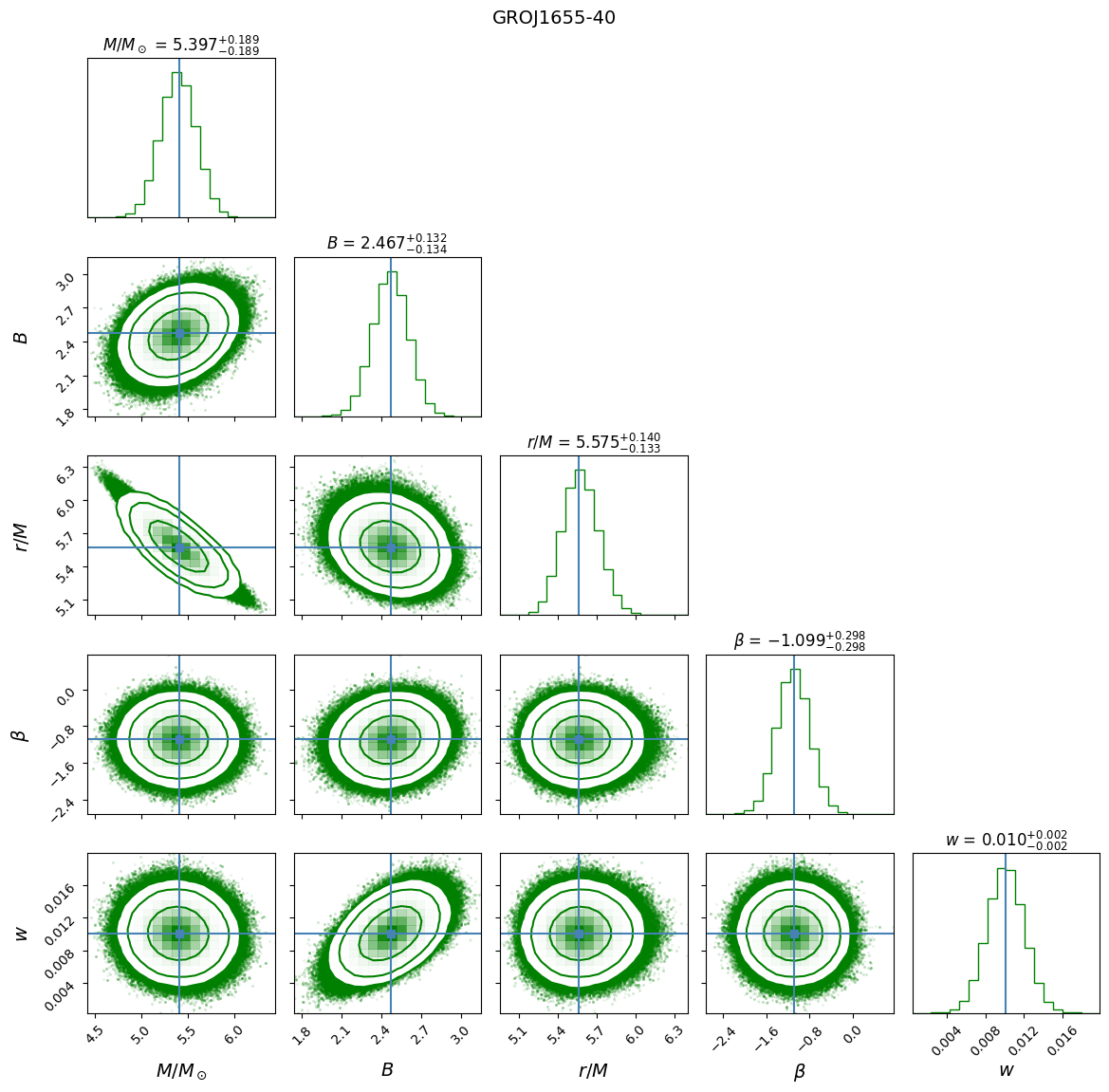}}
    \subfloat[]{\includegraphics[width=0.45\linewidth]{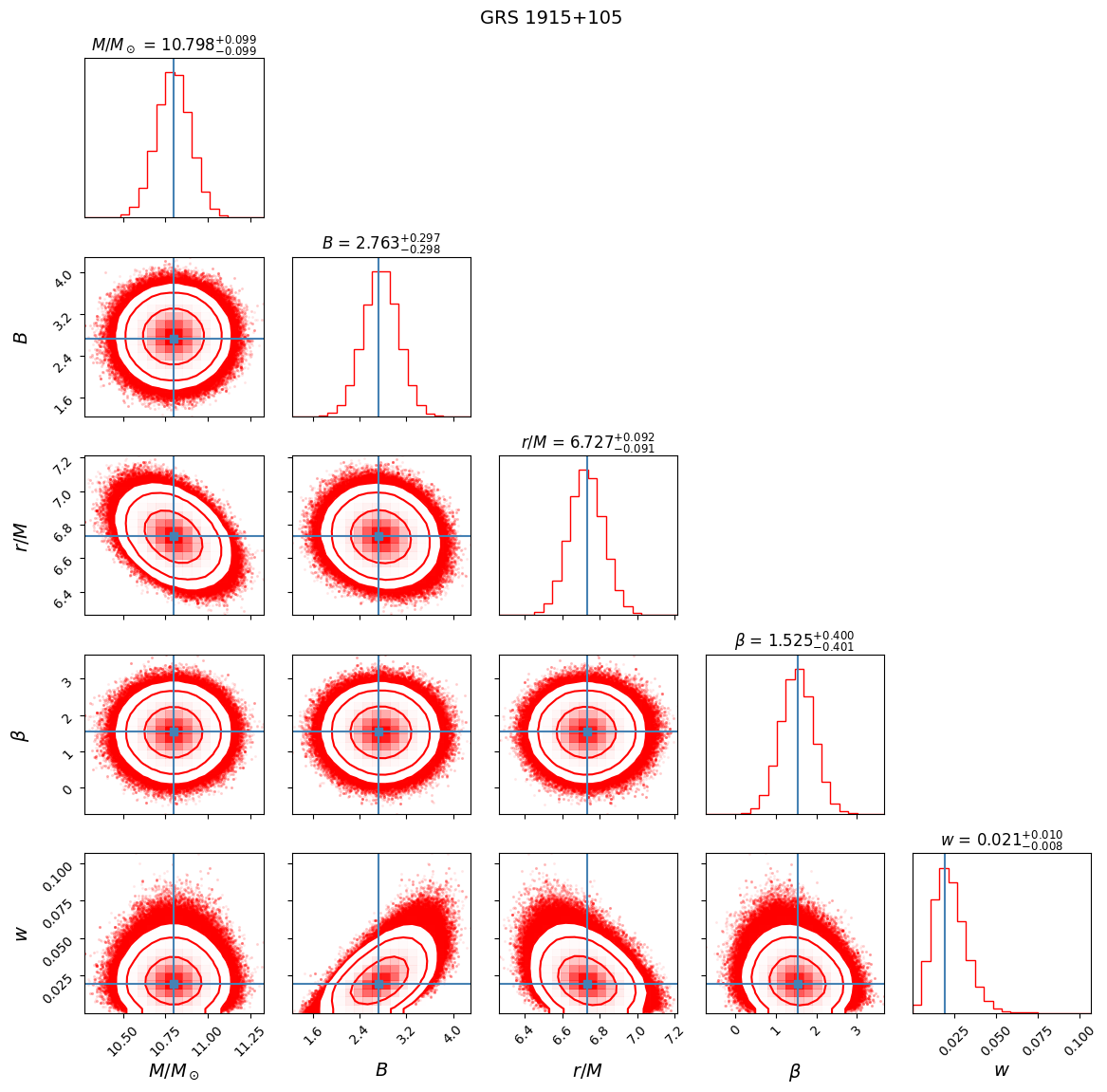}}\\
    \subfloat[]{\includegraphics[width=0.45\linewidth]{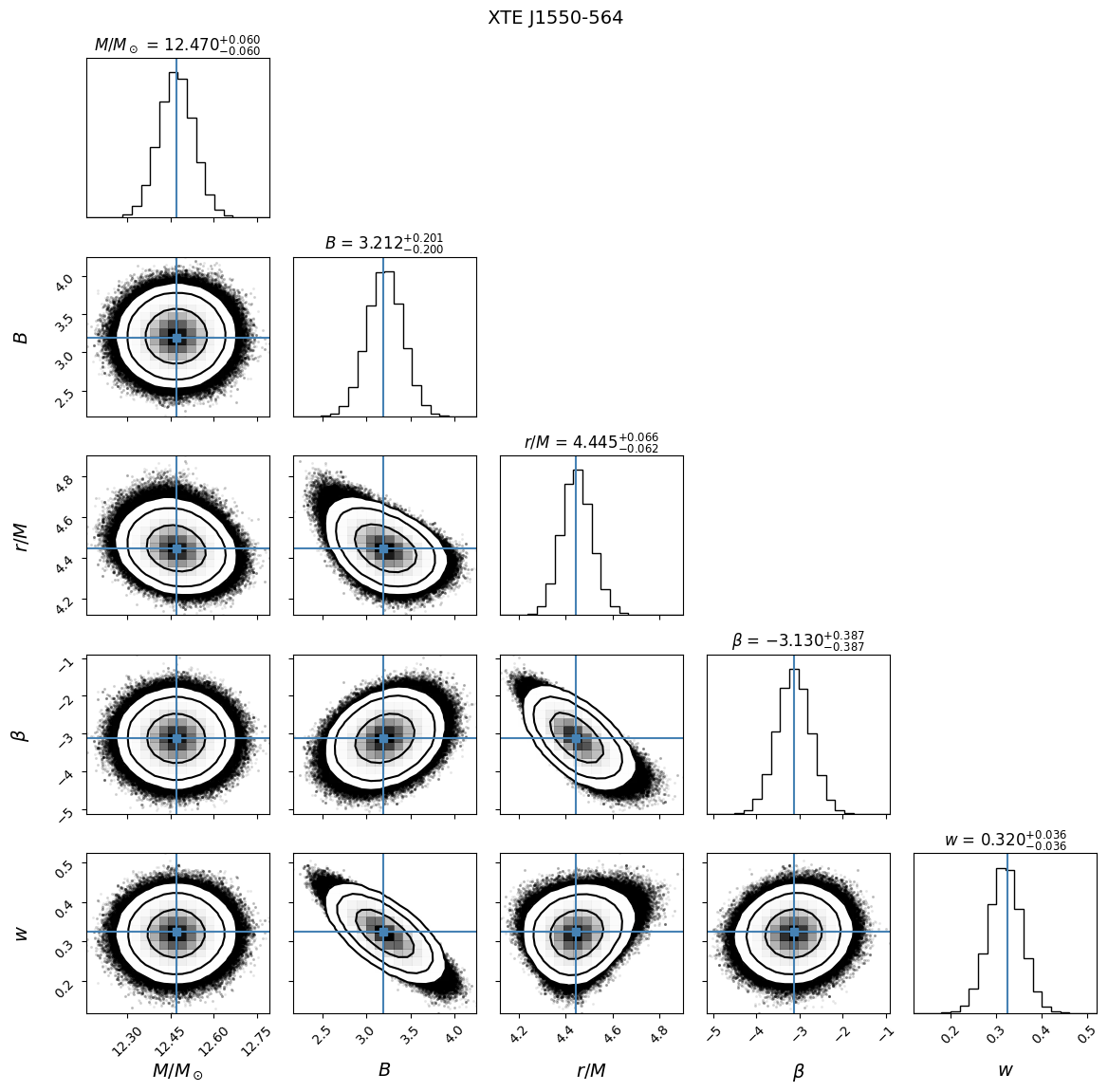}}
    \subfloat[]{\includegraphics[width=0.45\linewidth]{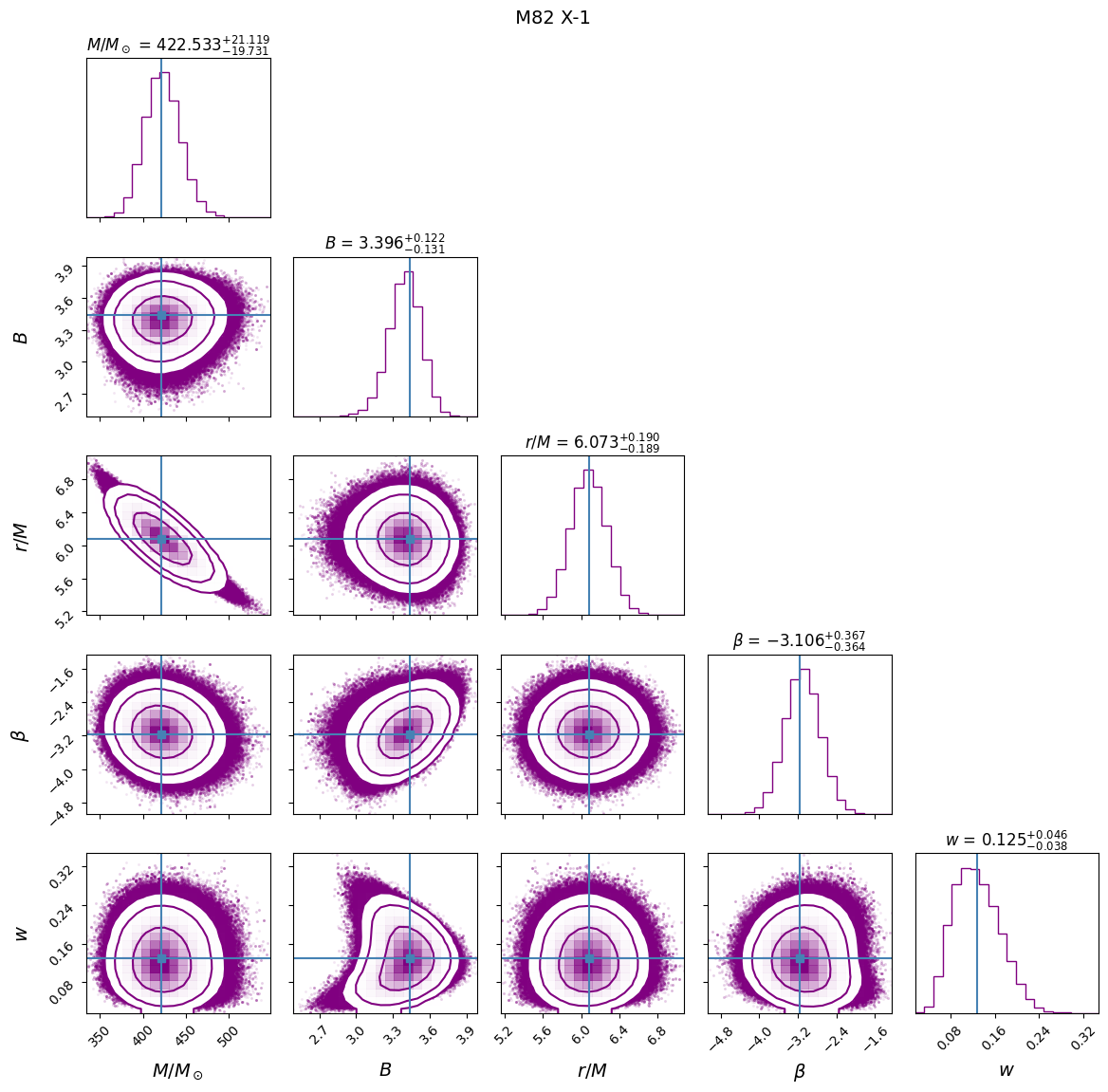}}\\
    \subfloat[]{\includegraphics[width=0.45\linewidth]{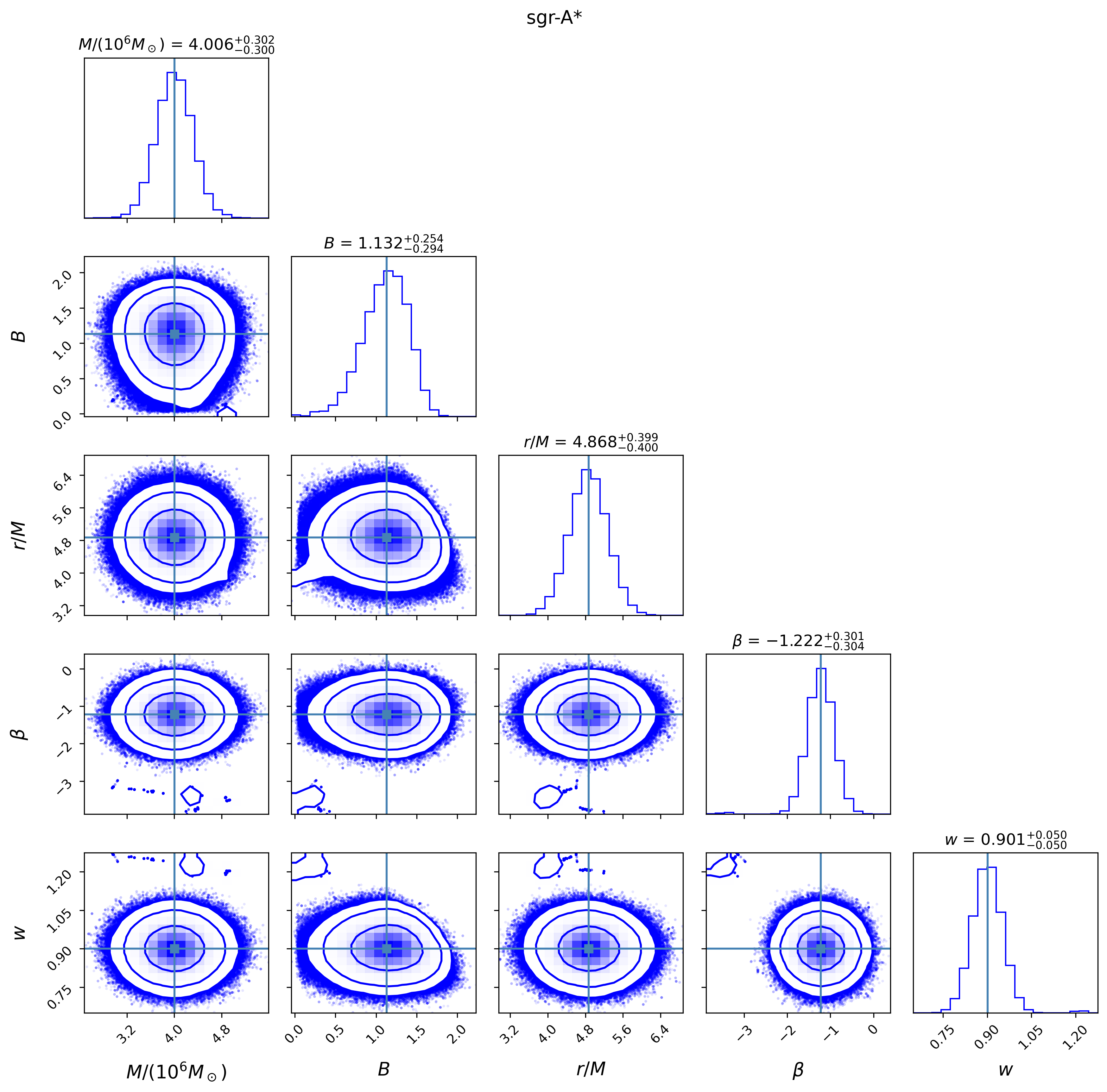}}
    \caption{\footnotesize{\it  Constraints on magnetized BH parameters of the QPO orbit from a five-dimensional MCMC analysis using the QPO data for the stellar mass BHs GRS 1915+105 ,GRO J1655-40 and XTE J1550-564 , intermediate-mass BH M82 X-1  and supermassive BH $\textrm{Sgr A}^\star$ and in the RP model.}}
    \label{MCMC plot}
\end{figure}

Fig.\ref{MCMC plot} displays the posterior probability distributions obtained from the five-dimensional MCMC analysis for all considered sources. The contour plots represent the $1\sigma$ (68\%), $2\sigma$ (95\%), and $3\sigma$ (99\%) confidence levels. The corresponding best-fit parameter values and uncertainties are reported in Table~\ref{postriors}.

\begin{table}[!ht]
\centering
\caption{\footnotesize{\it The best fit values of magnetized BH parameters.}}\label{postriors}
\adjustbox{max width=\textwidth}{
\begin{tabular}{lccccc}
\toprule
Parameter & {XTE J1550-564} & {GRO J1655-40} & {GRS 1915+105} & {Sgr \( A^* \)} & {M82 X-1} \\
\midrule
\vspace{3mm}
\(M/M_{\odot}\) & $12.470^{+0.060}_{-0.060}$ & $5.397^{+0.189}_{-0.189}$ & $10.798^{+0.099}_{-0.099}$ & $4.004^{+0.302}_{-0.301}$ & $422.533^{+21.119}_{-19.731}$ \\
\vspace{3mm}
\( \mathcal{B}\) & $3.212^{+0.201}_{-0.200}$ & $2.467^{+0.132}_{-0.134}$ & $2.763^{+0.297}_{-0.298}$ & $1.132^{+0.254}_{-0.295}$ & $3.396^{+0.122}_{-0.131}$ \\
\vspace{3mm}
\(r/M\) & $4.445^{+0.066}_{-0.062}$ & $5.575^{+0.140}_{-0.133}$ & $6.727^{+0.092}_{-0.091}$ & $4.873^{+0.394}_{-0.399}$ & $6.073^{+0.190}_{-0.183}$ \\
\vspace{3mm}
\(\mathcal{\beta}\) & $-3.130^{+0.387}_{-0.387}$ & $-1.099^{+0.298}_{-0.298}$ & $1.525^{+0.400}_{-0.401}$ & $-1.222^{+0.301}_{-0.304}$ & $-3.106^{+0.367}_{-0.364}$ \\
\(w\) & $0.320^{+0.036}_{-0.036}$ & $0.010^{+0.002}_{-0.002}$ & $0.021^{+0.010}_{-0.008}$ & $0.901^{+0.050}_{-0.050}$ & $0.125^{+0.046}_{-0.038}$ \\
\bottomrule
\end{tabular}
}
\end{table}
Among the five sources, $\textrm{Sgr A}^\star$ exhibits the weakest magnetic field strength, consistent with its low accretion activity. The magnetic field geometry parameter $w$, which controls the alignment of the parabolic magnetic field lines \cite{kolovs2023charged}, shows significant variation across different mass scales. Notably, GRO~J1655--40 and GRS~1915+105 display nearly identical magnetic field strengths and values of $w$. We recall that $w=0$ corresponds to a monopolar magnetic field configuration \cite{kolovs2023charged}, and both sources lie close to this regime.

\begin{figure}[H]
    \centering
    \subfloat[]{\includegraphics[width=0.45\linewidth]{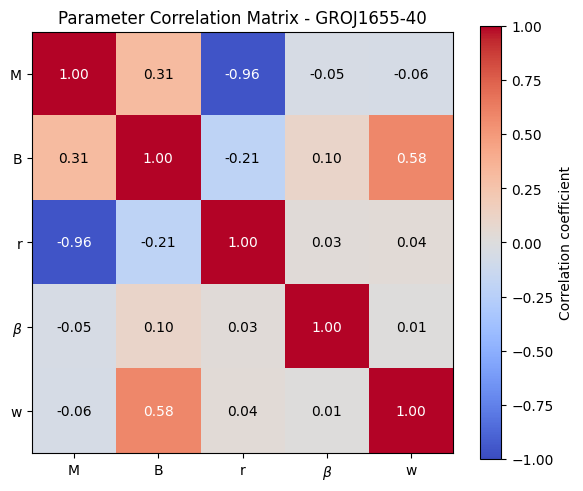}}
    \subfloat[]{\includegraphics[width=0.45\linewidth]{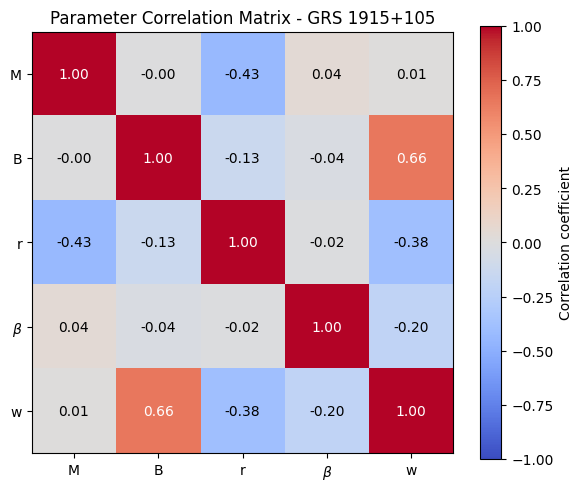}}\\
    \subfloat[]{\includegraphics[width=0.45\linewidth]{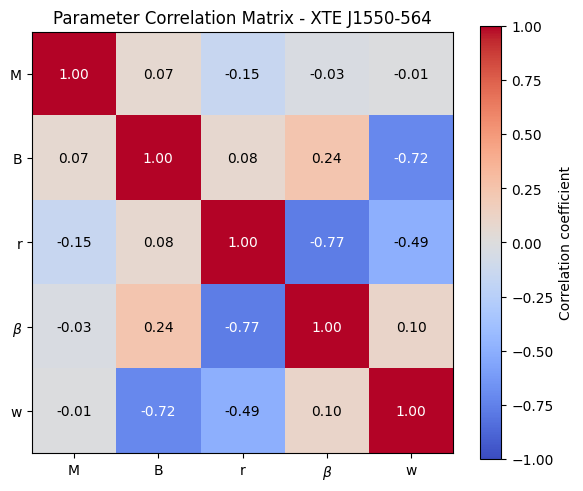}}
    \subfloat[]{\includegraphics[width=0.45\linewidth]{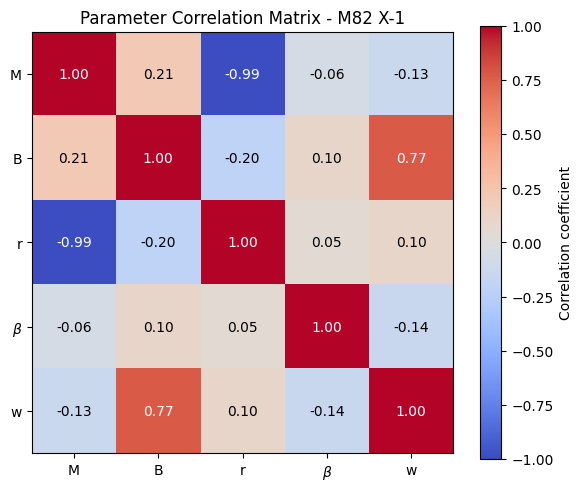}}\\
    \subfloat[]{\includegraphics[width=0.45\linewidth]{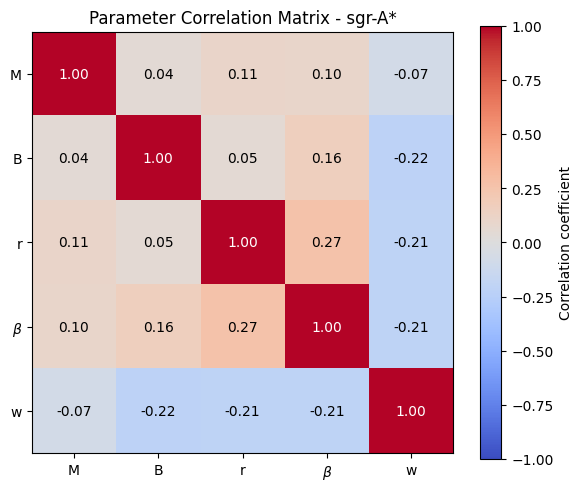}}
    \caption{\footnotesize{\it  Parameter correlation matrices derived from the MCMC posterior samples for the analyzed black hole sources, illustrating the level of degeneracy among the parameters $(M, B, r/M, \beta, w)$.}}
    \label{correlation matrix}
\end{figure}

Fig.\ref{correlation matrix} presents the parameter correlation
matrices derived from the MCMC posterior samples for all analyzed
black hole sources. 
These matrices are used to assess the
identifiability of the model parameters and to investigate possible
degeneracies arising from fitting the five–dimensional parameter
space $(M,B,r/M,\beta,w)$ to the two observed QPO frequencies
$(\nu_U,\nu_L)$.

A strong anti-correlation between the black hole mass $M$ and the
orbital radius $r$ is observed for several sources (e.g. $\rho\simeq
-0.96$ for GRO~J1655–40 and $\rho\simeq -0.99$ for M82~X–1). Such a
behavior is expected in QPO models since the characteristic
frequencies scale approximately as $\nu \propto M^{-1}f(r/M)$.
Therefore, the $M-r$ correlation reflects the intrinsic scaling of
the model rather than a pathological parameter degeneracy.

More importantly, the correlation coefficients between the magnetic
field strength $B$ and the dipole coupling parameter $\beta$ remain
small for all analyzed sources ($|\rho|\lesssim0.25$). This indicates
that these parameters influence the particle dynamics through
distinct physical mechanisms: $B$ controls the Lorentz force acting
on the charged particle, while $\beta$ governs the dipole field
interaction that modifies the effective potential.

A moderate correlation is instead observed between $B$ and the
magnetosphere geometry parameter $w$, which is physically expected
since both parameters affect the structure of the paraboloidal
magnetosphere. Finally, the dipole coupling parameter $\beta$
remains only weakly correlated with the other model parameters,
indicating that the dipole interaction introduces an independent
physical effect.

Overall, the correlation matrices demonstrate that the parameters
are statistically identifiable, and that the MCMC analysis does not
suffer from significant degeneracy or overfitting.

\section{Conclusion}\label{conclusion}

\paragraph{}In this work, we investigated the dynamics and observational implications of a magnetized black hole interacting with a charged particle endowed with a magnetic dipole moment. By combining analytical modeling with Bayesian inference techniques, we constrained a set of key physical parameters characterizing the system, namely the black hole mass $M$, the magnetic field strength $B$, the QPO orbital radius $r$, the magnetic coupling parameter $\beta$, and the magnetic field configuration parameter $w$.

We first analyzed the motion of a magnetized charged particle moving in the combined gravitational field of the black hole and an external paraboloidal magnetic field. The interaction between the particle’s intrinsic magnetic dipole moment and the external magnetic field introduces a nontrivial coupling that significantly alters the particle dynamics. In particular, the magnetic field strength $B$ and the coupling parameter $\beta$ exhibit opposite effects on the effective potential. As $B$ ($\beta$) varies from negative to positive values, the effective potential $V_{\text{eff}}$ decreases (increases), leading to distinct modifications of stable orbital configurations.

At the level of the ISCO, we showed that when $B$ and $\beta$ share the same polarity, the ISCO radius moves closer to the event horizon for negative values and shifts outward for positive values of both parameters. Moreover, the coupling parameter $\beta$ plays a crucial role in shaping the phase-space structure of particle trajectories. For sufficiently small values of $\beta$, a closed boundary emerges in phase space, leading to particle capture by the black hole. In the presence of strong magnetic fields, negative values of $\beta$ efficiently confine particles near the black hole, enhancing magnetic trapping effects.

We also examined the radiative properties of accretion disks composed of magnetized charged particles orbiting the black hole. Our analysis revealed that both the radiation flux and the disk temperature increase as the magnetic field strength $B$ decreases, while variations in the coupling parameter $\beta$ produce a qualitatively similar effect. Near the ISCO, the disk temperature reaches very high values and the radiation flux becomes strongly enhanced, indicating that magnetic and relativistic effects are dominant in the innermost regions of the disk.

Furthermore, the spatial extent of these hot and luminous regions depends sensitively on the parameters $B$ and $\beta$. For $B>0$ and $\beta>0$ ($B<0$ and $\beta<0$), the region of enhanced temperature and radiation flux tends to expand (contract). When both parameters take negative values, the increase in temperature and radiation flux is particularly pronounced. These results demonstrate that the presence of an external magnetic field can significantly boost the disk luminosity, with opposite field orientations producing stronger emission. In addition, negative values of the coupling parameter $\beta$ amplify the radiative output, whereas positive values tend to suppress it. At large radii, all models converge, confirming that magnetic effects are primarily relevant in the inner accretion disk. Since both disk radiation properties and QPO frequencies originate from particle dynamics near the ISCO, magnetic-field–induced changes in the thermal emission provide an independent support for the QPO-based interpretation of the system.

In the second part of this work, we applied the dynamics of magnetized charged particles to the study of quasi-periodic oscillations within the relativistic precession (RP) model. In this framework, the upper and lower HF QPO frequencies, $v_U$ and $v_L$, are constructed from the azimuthal and radial epicyclic frequencies, $v_\phi$ and $v_r$. We analyzed the dependence of these frequencies on the magnetic field strength $B$ and the coupling parameter $\beta$. Our results indicate that positive values of $B$ significantly enhance the QPO frequencies, while negative values reduce them. Conversely, positive values of $\beta$ suppress the frequencies, whereas negative values lead to stronger and more pronounced QPO signals.

Finally, we employed a Markov Chain Monte Carlo (MCMC) analysis to place quantitative constraints on the magnetic field strength, the magnetic coupling parameter, the inclination of the external magnetic field lines, the black hole mass, and the QPO orbital radii. The analysis was performed using twin-peak QPO observations from three stellar-mass black holes (XTE~J1550-564, GRO~J1655-40, and GRS~1915+105), the intermediate-mass black hole candidate M82~X-1, and the supermassive black hole Sgr~A*. The posterior distributions and confidence contours are presented in Fig.\ref{MCMC plot}, while the corresponding best-fit values are summarized in Table~\ref{postriors}. Moreover, the parameter correlation matrices provide insight into whether the inferred parameters are correlated or independently constrained. A clear anti-correlation is observed between $M$ and $r$ for several sources, such as GRO~J1655–40 and M82~X-1, which is expected since QPO frequencies scale approximately as $\nu \propto M^{-1} f(r/M)$. Since our model focuses on the effects of an external paraboloidal magnetic field, we also examine the correlation between the magnetic field strength $B$ and the dipole coupling parameter $\beta$. We find that this correlation remains weak for all analyzed sources, indicating that the magnetic field strength and dipole interaction affect the QPO frequencies through distinct physical mechanisms and can therefore be constrained independently.

These results demonstrate that QPO observations provide a powerful probe of magnetic environments around black holes and offer a consistent framework for constraining both gravitational and electromagnetic properties across a wide range of black hole masses.

Future work will extend this analysis to rotating black holes and more realistic magnetic field configurations inspired by GRMHD simulations. Investigating chaotic dynamics, polarization signatures, and multi-frequency QPO correlations may provide additional observational tests. Combining QPO modeling with black hole shadow and X-ray polarimetric observations offers a promising avenue to probe magnetic effects in the strong-gravity regime.

\section*{Acknowledgements}
\paragraph{}H. El M would like to acknowledge the networking support of the COST Action 
 CA 22113 - Fundamental challenges in theoretical physics (Theory and Challenges), 
CA 21136 - Addressing observational tensions in cosmology with systematics and fundamental physics (CosmoVerse), and 
CA 23130 - Bridging high and low energies in search of quantum gravity (BridgeQG). He also thanks IOP for its support.

Z. A expresses gratitude for the financial support he receives from the National Center for Scientific and Technical Research (CNRST) of Morocco, within the framework of the PhD-Associate Scholarship – PASS program.

This work was carried out under the project UIZ 2025 Scientific Research Projects: {\tt PRJ-2025-81}.

\section*{CRediT authorship contribution statement}
{\bf Z. Ahal}: Writing – original draft, Visualization, Software, Investigation.\\
{\bf H. El Moumni}: Writing – original draft, Supervision, Methodology, Conceptualization, Funding acquisition.\\
{\bf K. Masmar}: Writing – review \& editing, Supervision, Project administration, Formal analysis.

\section*{Declaration of competing interest}
The authors declare that they have no known competing financial interests or personal relationships that could have appeared to influence the work reported in this paper.

\bibliographystyle{unsrt}
\bibliography{main}

@article{EventHorizonTelescope:2024hpu,
    author = "Akiyama, Kazunori and others",
    collaboration = "Event Horizon Telescope",
    title = "{First Sagittarius A* Event Horizon Telescope Results. VII. Polarization of the Ring}",
    doi = "10.3847/2041-8213/ad2df0",
    journal = "Astrophys. J. Lett.",
    volume = "964",
    number = "2",
    pages = "L25",
    year = "2024"
}

@article{EventHorizonTelescope:2021bee,
    author = "Akiyama, Kazunori and others",
    collaboration = "Event Horizon Telescope",
    title = "{First M87 Event Horizon Telescope Results. VII. Polarization of the Ring}",
    eprint = "2105.01169",
    archivePrefix = "arXiv",
    primaryClass = "astro-ph.HE",
    reportNumber = "FERMILAB-PUB-21-849-PPD",
    doi = "10.3847/2041-8213/abe71d",
    journal = "Astrophys. J. Lett.",
    volume = "910",
    number = "1",
    pages = "L12",
    year = "2021"
}

@article{crinquand2021synthetic,
  title={Synthetic gamma-ray light curves of Kerr black hole magnetospheric activity from particle-in-cell simulations},
  author={Crinquand, Benjamin and Cerutti, Beno{\^\i}t and Dubus, Guillaume and Parfrey, Kyle and Philippov, Alexander},
  journal={Astronomy \& Astrophysics},
  volume={650},
  pages={A163},
  year={2021},
  publisher={EDP Sciences}
}

@article{mckinney2007disc,
  title={Disc--jet coupling in black hole accretion systems--I. General relativistic magnetohydrodynamical models},
  author={McKinney, Jonathan C and Narayan, Ramesh},
  journal={Monthly Notices of the Royal Astronomical Society},
  volume={375},
  number={2},
  pages={513--530},
  year={2007},
  publisher={Blackwell Publishing Ltd Oxford, UK}
}

@article{blandford1977electromagnetic,
  title={Electromagnetic extraction of energy from Kerr black holes},
  author={Blandford, Roger D and Znajek, Roman L},
  journal={Monthly Notices of the Royal Astronomical Society},
  volume={179},
  number={3},
  pages={433--456},
  year={1977},
  publisher={Oxford University Press Oxford, UK}
}

@article{nakamura2018parabolic,
  title={Parabolic jets from the spinning black hole in M87},
  author={Nakamura, Masanori and Asada, Keiichi and Hada, Kazuhiro and Pu, Hung-Yi and Noble, Scott and Tseng, Chihyin and Toma, Kenji and Kino, Motoki and Nagai, Hiroshi and Takahashi, Kazuya and others},
  journal={The Astrophysical Journal},
  volume={868},
  number={2},
  pages={146},
  year={2018},
  publisher={IOP Publishing}
}

@article{jumaniyozov2024collisions,
  title={Collisions and dynamics of particles with magnetic dipole moment and electric charge near magnetized rotating Kerr black holes},
  author={Jumaniyozov, Shokhzod and Khan, Saeed Ullah and Rayimbaev, Javlon and Abdujabbarov, Ahmadjon and Ahmedov, Bobomurat},
  journal={The European Physical Journal C},
  volume={84},
  number={3},
  pages={291},
  year={2024},
  publisher={Springer}
}

@article{murodov2023dynamics,
  title={Dynamics of particles with electric charge and magnetic dipole moment near Schwarzschild-MOG black hole},
  author={Murodov, Sardor and Rayimbaev, Javlon and Ahmedov, Bobomurat and Hakimov, Abdullo},
  journal={Symmetry},
  volume={15},
  number={11},
  pages={2084},
  year={2023},
  publisher={MDPI}
}

@article{rayimbaev2024particles,
  title={Particles with magnetic dipole moment orbiting magnetized Schwarzschild black holes: Applications to orbits of hot-spots around Sgr A},
  author={Rayimbaev, Javlon and Ahmedov, Bobomurat and Stuchlik, Zdenek},
  journal={Physics of the Dark Universe},
  volume={45},
  pages={101516},
  year={2024},
  publisher={Elsevier}
}

@article{preti2004general,
  title={General relativistic dynamics of polarized particles in electromagnetic fields},
  author={Preti, Giovanni},
  journal={Physical Review D},
  volume={70},
  number={2},
  pages={024012},
  year={2004},
  publisher={APS}
}

@article{stuchlik2019magnetized,
  title={Magnetized black holes: ionized keplerian disks and acceleration of ultra-high energy particles},
  author={Stuchl{\'\i}k, Zden{\v{e}}k and Kolo{\v{s}}, Martin and Tursunov, Arman},
  journal={Multidisciplinary Digital Publishing Institute Proceedings},
  volume={17},
  number={1},
  pages={13},
  year={2019}
}

@article{kolovs2020simulations,
  title={Simulations of black hole accretion torus in various magnetic field configurations},
  author={Kolo{\v{s}}, Martin and Janiuk, Agnieszka},
  journal={arXiv preprint arXiv:2004.07535},
  year={2020}
}

@article{stella1999correlations,
  title={Correlations in the quasi-periodic oscillation frequencies of low-mass X-ray binaries and the relativistic precession model},
  author={Stella, Luigi and Vietri, Mario and Morsink, Sharon M},
  journal={The Astrophysical Journal},
  volume={524},
  number={1},
  pages={L63},
  year={1999},
  publisher={IOP Publishing}
}

@article{hazarika2025signatures,
  title={Signatures of NED on Quasi periodic Oscillations of a Magnetically Charged Black Hole},
  author={Hazarika, Bidyut and Gohain, Mrinnoy M and Phukon, Prabwal},
  journal={arXiv preprint arXiv:2504.07821},
  year={2025}
}

@article{Abramowicz:2001bi,
    author = "Abramowicz, Marek Artur and Kluzniak, Wlodek",
    title = "{A Precise determination of angular momentum in the black hole candidate GRO J1655-40}",
    eprint = "astro-ph/0105077",
    archivePrefix = "arXiv",
    doi = "10.1051/0004-6361:20010791",
    journal = "Astron. Astrophys.",
    volume = "374",
    pages = "L19",
    year = "2001"
}

@article{remillard2002evidence,
  title={Evidence for harmonic relationships in the high-frequency quasi-periodic oscillations of XTE J1550--564 and GRO J1655--40},
  author={Remillard, Ronald A and Muno, Michael P and McClintock, Jeffrey E and Orosz, Jerome A},
  journal={The Astrophysical Journal},
  volume={580},
  number={2},
  pages={1030},
  year={2002},
  publisher={IOP Publishing}
}

@article{morgan1997rxte,
  title={RXTE observations of QPOs in the black hole candidate GRS 1915+ 105},
  author={Morgan, EH and Remillard, RA and Greiner, J},
  journal={The Astrophysical Journal},
  volume={482},
  number={2},
  pages={993},
  year={1997},
  publisher={IOP Publishing}
}

@article{strohmayer2001discovery,
  title={Discovery of a 450 hz quasi-periodic oscillation from the microquasar gro j1655--40 with the rossi x-ray timing explorer},
  author={Strohmayer, Tod E},
  journal={The Astrophysical Journal},
  volume={552},
  number={1},
  pages={L49},
  year={2001},
  publisher={IOP Publishing}
}

@incollection{van1989fourier,
  title={Fourier techniques in X-ray timing},
  author={Van der Klis, M},
  booktitle={Timing neutron stars},
  pages={27--69},
  year={1989},
  publisher={Springer}
}

@article{fiorito2004m82,
  title={Is M82 X-1 really an intermediate-mass black hole? X-ray spectral and timing evidence},
  author={Fiorito, Ralph and Titarchuk, Lev},
  journal={The Astrophysical Journal},
  volume={614},
  number={2},
  pages={L113},
  year={2004},
  publisher={IOP Publishing}
}

@article{stuchlik2015mass,
  title={Mass of intermediate black hole in the source M82 X-1 restricted by models of twin high-frequency quasi-periodic oscillations},
  author={Stuchl{\'\i}k, Zden{\v{e}}k and Kolo{\v{s}}, Martin},
  journal={Monthly Notices of the Royal Astronomical Society},
  volume={451},
  number={3},
  pages={2575--2588},
  year={2015},
  publisher={Oxford University Press}
}

@article{foreman2013emcee,
  title={emcee: the MCMC hammer},
  author={Foreman-Mackey, Daniel and Hogg, David W and Lang, Dustin and Goodman, Jonathan},
  journal={Publications of the Astronomical Society of the Pacific},
  volume={125},
  number={925},
  pages={306},
  year={2013},
  publisher={IOP Publishing}
}

@inproceedings{gendreau2016neutron,
  title={The neutron star interior composition explorer (NICER): design and development},
  author={Gendreau, Keith C and Arzoumanian, Zaven and Adkins, Phillip W and Albert, Cheryl L and Anders, John F and Aylward, Andrew T and Baker, Charles L and Balsamo, Erin R and Bamford, William A and Benegalrao, Suyog S and others},
  booktitle={Space telescopes and instrumentation 2016: Ultraviolet to gamma ray},
  volume={9905},
  pages={420--435},
  year={2016},
  organization={SPIE}
}

@article{torok2005orbital,
  title={The orbital resonance model for twin peak kHz quasi periodic oscillations in microquasars},
  author={T{\"o}r{\"o}k, Gabriel and Abramowicz, Marek A and Klu{\'z}niak, W and Stuchl{\'\i}k, Z},
  journal={Astronomy \& Astrophysics},
  volume={436},
  number={1},
  pages={1--8},
  year={2005},
  publisher={EDP Sciences}
}

@article{Motta:2013wwa,
    author = "Motta, S. E. and Mu\~noz-Darias, T. and Sanna, A. and Fender, R. and Belloni, T. and Stella, L.",
    title = "{Black hole spin measurements through the relativistic precession model: XTE J1550-564}",
    eprint = "1312.3114",
    archivePrefix = "arXiv",
    primaryClass = "astro-ph.HE",
    doi = "10.1093/mnrasl/slt181",
    journal = "Mon. Not. Roy. Astron. Soc.",
    volume = "439",
    pages = "65",
    year = "2014"
}

@article{nishonov2025qpos,
  title={QPOs from charged particles around charged black holes in STVG},
  author={Nishonov, Isomiddin and Murodov, Sardor and Ahmedov, Bobomurat and Khan, Saeed Ullah and Rayimbaev, Javlon and Ibragimov, Inomjon and Sabirov, Sardor},
  journal={The European Physical Journal C},
  volume={85},
  number={9},
  pages={1029},
  year={2025},
  publisher={Springer}
}

@article{Askour:2024nky,
    author = "Askour, N. and Belhaj, A. and Chakhchi, L. and El Moumni, H. and Masmar, K.",
    title = "{On M87{\textasteriskcentered} and SgrA{\textasteriskcentered} observational constraints of Dunkl black holes}",
    eprint = "2412.09196",
    archivePrefix = "arXiv",
    primaryClass = "gr-qc",
    doi = "10.1016/j.jheap.2025.100349",
    journal = "JHEAp",
    volume = "46",
    pages = "100349",
    year = "2025"
}

@article{Papapetrou:1951pa,
    author = "Papapetrou, Achille",
    title = "{Spinning test particles in general relativity. 1.}",
    doi = "10.1098/rspa.1951.0200",
    journal = "Proc. Roy. Soc. Lond. A",
    volume = "209",
    pages = "248--258",
    year = "1951"
}

@article{Chakhchi:2024tzo,
    author = "Chakhchi, L. and El Moumni, H. and Masmar, K.",
    title = "{Signatures of the accelerating black holes with a cosmological constant from the Sgr A{\ensuremath{\star}} and M87{\ensuremath{\star}} shadow prospects}",
    eprint = "2403.09756",
    archivePrefix = "arXiv",
    primaryClass = "gr-qc",
    doi = "10.1016/j.dark.2024.101501",
    journal = "Phys. Dark Univ.",
    volume = "44",
    pages = "101501",
    year = "2024"
}

@article{ahal2025modeling,
  title={Modeling HF-QPOs in microquasars and AGNs: charged particles around black holes with CDM halos},
  author={Ahal, Zakaria and El Moumni, H and Masmar, Karima},
  journal={The European Physical Journal C},
  volume={85},
  number={10},
  pages={1--36},
  year={2025},
  publisher={Springer}
}

@article{abramowicz2004orbital,
  title={The orbital resonance model for twin peak kHz QPOs},
  author={Abramowicz, Marek A and Kluzniak, Wlodek and Stuchlik, Zdenek and Torok, Gabriel},
  journal={arXiv preprint astro-ph/0401464},
  year={2004}
}

@article{smith2021confrontation,
  title={Confrontation of observation and theory: high-frequency QPOs in X-ray binaries, tidal disruption events, and active galactic nuclei},
  author={Smith, Krista Lynne and Tandon, Celia R and Wagoner, Robert V},
  journal={The Astrophysical Journal},
  volume={906},
  number={2},
  pages={92},
  year={2021},
  publisher={IOP Publishing}
}

@article{tremaine2014dynamics,
  title={Dynamics of warped accretion discs},
  author={Tremaine, Scott and Davis, Shane W},
  journal={Monthly Notices of the Royal Astronomical Society},
  volume={441},
  number={2},
  pages={1408--1434},
  year={2014},
  publisher={Oxford University Press}
}

@article{borah2025black,
  title={Black hole quasi-periodic oscillations in the presence of Gauss-Bonnet trace anomaly},
  author={Borah, Rupam Jyoti and Goswami, Umananda Dev},
  journal={Physics Letters B},
  pages={140124},
  year={2025},
  publisher={Elsevier}
}

@article{ripperda2022black,
  title={Black hole flares: ejection of accreted magnetic flux through 3D plasmoid-mediated reconnection},
  author={Ripperda, Bart and Liska, Matthew and Chatterjee, Koushik and Musoke, Gibwa and Philippov, Alexander A and Markoff, Sera B and Tchekhovskoy, Alexander and Younsi, Ziri},
  journal={The Astrophysical Journal Letters},
  volume={924},
  number={2},
  pages={L32},
  year={2022},
  publisher={IOP Publishing}
}

@article{narzilloev2021dynamics,
  title={Dynamics of charged particles and magnetic dipoles around magnetized quasi-Schwarzschild black holes},
  author={Narzilloev, Bakhtiyor and Rayimbaev, Javlon and Abdujabbarov, Ahmadjon and Ahmedov, Bobomurat and Bambi, Cosimo},
  journal={The European Physical Journal C},
  volume={81},
  number={3},
  pages={269},
  year={2021},
  publisher={Springer}
}

@article{tchekhovskoy2010black,
  title={Black hole spin and the radio loud/quiet dichotomy of active galactic nuclei},
  author={Tchekhovskoy, Alexander and Narayan, Ramesh and McKinney, Jonathan C},
  journal={The Astrophysical Journal},
  volume={711},
  number={1},
  pages={50},
  year={2010},
  publisher={IOP Publishing}
}

@article{kolovs2023charged,
  title={Charged particle dynamics in parabolic magnetosphere around Schwarzschild black hole},
  author={Kolo{\v{s}}, Martin and Shahzadi, Misbah and Tursunov, Arman},
  journal={The European Physical Journal C},
  volume={83},
  number={4},
  pages={323},
  year={2023},
  publisher={Springer}
}

@article{wald1974black,
  title={Black hole in a uniform magnetic field},
  author={Wald, Robert M},
  journal={Physical Review D},
  volume={10},
  number={6},
  pages={1680},
  year={1974},
  publisher={APS}
}

@article{piotrovich2010magnetic,
  title={Magnetic fields of black holes and the variability plane},
  author={Piotrovich, M Yu and Silant'ev, NA and Gnedin, Yu N and Natsvlishvili, TM},
  journal={arXiv preprint arXiv:1002.4948},
  year={2010}
}

@article{bouchy2001p,
  title={P-mode observations on Cen A},
  author={Bouchy, Francois and Carrier, Fabien},
  journal={Astronomy \& Astrophysics},
  volume={374},
  number={1},
  pages={L5--L8},
  year={2001},
  publisher={EDP Sciences}
}

@article{shermatov2025qpos,
  title={QPOs analyses and circular orbits of charged particles around magnetized black holes in Bertotti--Robinson geometry},
  author={Shermatov, Abubakir and Rayimbaev, Javlon and L{\"u}tf{\"u}olu, Bekir Can and Abdujabbarov, Ahmadjon and Sardor, Sabirov and Ibragimov, Inomjon and Vapayev, Murodbek and Kuyliev, Bahrom},
  journal={The European Physical Journal C},
  volume={85},
  number={9},
  pages={1017},
  year={2025},
  publisher={Springer}
}

@article{shahzadi2021epicyclic,
  title={Epicyclic oscillations in spinning particle motion around Kerr black hole applied in models fitting the quasi-periodic oscillations observed in microquasars and AGNs},
  author={Shahzadi, Misbah and Kolo{\v{s}}, Martin and Stuchl{\'\i}k, Zden{\v{e}}k and Habib, Yousaf},
  journal={The European Physical Journal C},
  volume={81},
  number={12},
  pages={1067},
  year={2021},
  publisher={Springer}
}

@article{jumaniyozov2024circular,
  title={Circular motion and QPOs near black holes in Kalb--Ramond gravity},
  author={Jumaniyozov, Shokhzod and Khan, Saeed Ullah and Rayimbaev, Javlon and Abdujabbarov, Ahmadjon and Urinbaev, Sharofiddin and Murodov, Sardor},
  journal={The European Physical Journal C},
  volume={84},
  number={9},
  pages={964},
  year={2024},
  publisher={Springer}
}

@article{donmez2024perturbing,
  title={Perturbing the Stable Accretion Disk in Kerr and 4D Einstein--Gauss--Bonnet Gravities: Comprehensive Analysis of Instabilities and Dynamics},
  author={Donmez, Orhan},
  journal={Research in Astronomy and Astrophysics},
  volume={24},
  number={8},
  pages={085001},
  year={2024},
  publisher={IOP Publishing}
}

@article{donmez2024proposing,
  title={Proposing a physical mechanism to explain various observed sources of QPOs by simulating the dynamics of accretion disks around the black holes},
  author={Donmez, Orhan},
  journal={The European Physical Journal C},
  volume={84},
  number={5},
  pages={524},
  year={2024},
  publisher={Springer}
}

@article{donmez2024comparison,
  title={The comparison of alternative spacetimes using the spherical accretion around the black hole},
  author={Donmez, Orhan},
  journal={Modern Physics Letters A},
  volume={39},
  number={16},
  pages={2450076},
  year={2024},
  publisher={World Scientific}
}

@article{remillard2006x,
  title={X-ray properties of black-hole binaries},
  author={Remillard, Ronald A and McClintock, Jeffrey E},
  journal={Annu. Rev. Astron. Astrophys.},
  volume={44},
  number={1},
  pages={49--92},
  year={2006},
  publisher={Annual Reviews}
}

@article{rayimbaev2022radio,
  title={Radio Pulsars in an Electromagnetic Universe},
  author={Rayimbaev, Javlon and Jumaniyozov, Shokhzod and Umaraliyev, Maksud and Abdujabbarov, Ahmadjon},
  journal={Universe},
  volume={8},
  number={10},
  pages={496},
  year={2022},
  publisher={MDPI}
}

@article{abramowicz2001precise,
  title={A precise determination of black hole spin in GRO J1655-40},
  author={Abramowicz, Marek Artur and Klu{\'z}niak, W},
  journal={Astronomy \& Astrophysics},
  volume={374},
  number={3},
  pages={L19--L20},
  year={2001},
  publisher={EDP Sciences}
}

@article{mckee2007theory,
  title={Theory of star formation},
  author={McKee, Christopher F and Ostriker, Eve C},
  journal={Annu. Rev. Astron. Astrophys.},
  volume={45},
  number={1},
  pages={565--687},
  year={2007},
  publisher={Annual Reviews}
}

@article{blandford2000acceleration,
  title={Acceleration of ultra high energy cosmic rays},
  author={Blandford, RD},
  journal={Physica Scripta},
  volume={2000},
  number={T85},
  pages={191},
  year={2000},
  publisher={IOP Publishing}
}

@article{fraschetti2008acceleration,
  title={On the acceleration of ultra-high-energy cosmic rays},
  author={Fraschetti, Federico},
  journal={Philosophical Transactions of the Royal Society A: Mathematical, Physical and Engineering Sciences},
  volume={366},
  number={1884},
  pages={4417--4428},
  year={2008},
  publisher={The Royal Society London}
}

@article{tursunov2020supermassive,
  title={Supermassive black holes as possible sources of ultrahigh-energy cosmic rays},
  author={Tursunov, Arman and Stuchl{\'\i}k, Zden{\v{e}}k and Kolo{\v{s}}, Martin and Dadhich, Naresh and Ahmedov, Bobomurat},
  journal={The Astrophysical Journal},
  volume={895},
  number={1},
  pages={14},
  year={2020},
  publisher={IOP Publishing}
}

@article{koide2002extraction,
  title={Extraction of black hole rotational energy by a magnetic field and the formation of relativistic jets},
  author={Koide, Shinji and Shibata, Kazunari and Kudoh, Takahiro and Meier, David L},
  journal={Science},
  volume={295},
  number={5560},
  pages={1688--1691},
  year={2002},
  publisher={American Association for the Advancement of Science}
}

@article{marti2015strong,
  title={A strong magnetic field in the jet base of a supermassive black hole},
  author={Marti-Vidal, Ivan and Muller, Sebastien and Vlemmings, Wouter and Horellou, Cathy and Aalto, Susanne},
  journal={Science},
  volume={348},
  number={6232},
  pages={311--314},
  year={2015},
  publisher={American Association for the Advancement of Science}
}

@book{biskamp2003magnetohydrodynamic,
  title={Magnetohydrodynamic turbulence},
  author={Biskamp, Dieter},
  year={2003},
  publisher={Cambridge University Press}
}

@article{yamada2010magnetic,
  title={Magnetic reconnection},
  author={Yamada, Masaaki and Kulsrud, Russell and Ji, Hantao},
  journal={Reviews of modern physics},
  volume={82},
  number={1},
  pages={603--664},
  year={2010},
  publisher={APS}
}

@article{kolovs2019charged,
  title={Charged particle motion around Schwarzschild black hole with split monopole magnetosphere},
  author={Kolo{\v{s}}, Martin and Bardiev, Dilshodbek and Juraev, Bakhtinur},
  journal={Proceedings of RAGtime},
  pages={20--21},
  year={2019}
}

@article{kolovs2025charged,
  title={Charged particle dynamics in magnetosphere generated by current loop around Schwarzschild black hole},
  author={Kolo{\v{s}}, Martin and Kofro{\v{n}}, David},
  journal={arXiv preprint arXiv:2509.11518},
  year={2025}
}

@article{kolovs2015quasi,
  title={Quasi-harmonic oscillatory motion of charged particles around a Schwarzschild black hole immersed in a uniform magnetic field},
  author={Kolo{\v{s}}, Martin and Stuchl{\'\i}k, Zden{\v{e}}k and Tursunov, Arman},
  journal={Classical and Quantum Gravity},
  volume={32},
  number={16},
  pages={165009},
  year={2015},
  publisher={IOP Publishing}
}

@article{silvers2008magnetic,
  title={Magnetic fields in astrophysical objects},
  author={Silvers, LJ},
  journal={Philosophical Transactions of the Royal Society A: Mathematical, Physical and Engineering Sciences},
  volume={366},
  number={1884},
  pages={4453--4464},
  year={2008},
  publisher={The Royal Society London}
}

@article{wielebinski1993magnetic,
  title={Magnetic fields in galaxies},
  author={Wielebinski, Richard and Krause, F},
  journal={The Astronomy and Astrophysics Review},
  volume={4},
  number={4},
  pages={449--485},
  year={1993},
  publisher={Springer}
}

@article{goldreich1969pulsar,
  title={Pulsar electrodynamics},
  author={Goldreich, Peter and Julian, William H},
  journal={Astrophysical Journal, vol. 157, p. 869},
  volume={157},
  pages={869},
  year={1969}
}

@article{beck2016magnetic,
  title={Magnetic fields in spiral galaxies},
  author={Beck, Rainer},
  journal={The Astronomy and Astrophysics Review},
  volume={24},
  number={1},
  pages={4},
  year={2016},
  publisher={Springer}
}

@article{zaitsev2015particle,
  title={Particle acceleration and plasma heating in the chromosphere},
  author={Zaitsev, VV and Stepanov, AV},
  journal={Solar Physics},
  volume={290},
  number={12},
  pages={3559--3572},
  year={2015},
  publisher={Springer}
}

@article{abdujabbarov2014magnetized,
  title={Magnetized particle motion and acceleration around a Schwarzschild black hole in a magnetic field},
  author={Abdujabbarov, Ahmadjon and Ahmedov, Bobomurat and Rahimov, Ozodbek and Salikhbaev, Umar},
  journal={Physica Scripta},
  volume={89},
  number={8},
  pages={084008},
  year={2014},
  publisher={IOP Publishing}
}

@article{crutcher2012magnetic,
  title={Magnetic fields in molecular clouds},
  author={Crutcher, Richard M},
  journal={Annual Review of Astronomy and Astrophysics},
  volume={50},
  number={1},
  pages={29--63},
  year={2012},
  publisher={Annual Reviews}
}

@article{jumaniyozov2025black,
  title={Black holes surrounded by PFDM in Kalb-Ramond gravity: from thermodynamics to QPO tests},
  author={Jumaniyozov, Shokhzod and Murodov, Sardor and Rayimbaev, Javlon and Ibragimov, Inomjon and Madaminov, Bekzod and Urinbaev, Sharofiddin and Abdujabbarov, Ahmadjon},
  journal={The European Physical Journal C},
  volume={85},
  number={7},
  pages={797},
  year={2025},
  publisher={Springer}
}

@article{petterson1974magnetic,
  title={Magnetic field of a current loop around a Schwarzschild black hole},
  author={Petterson, Jacobus A},
  journal={Physical Review D},
  volume={10},
  number={10},
  pages={3166},
  year={1974},
  publisher={APS}
}

@article{vrba2025charged,
  title={Charged particles in dipole magnetosphere of neutron stars: epicyclic oscillations in and off-equatorial plane},
  author={Vrba, Jaroslav and Kolo{\v{s}}, Martin and Stuchl{\'\i}k, Zden{\v{e}}k},
  journal={The European Physical Journal Plus},
  volume={140},
  number={2},
  pages={1--25},
  year={2025},
  publisher={Springer}
}

@article{zhang2025curled,
  title={Curled orbit and epicyclic oscillation of charged particles around the weakly magnetized black hole in the presence of Lorentz violation},
  author={Zhang, Hai-Yang and Hu, Ya-Peng and An, Yu-Sen},
  journal={The European Physical Journal C},
  volume={85},
  number={7},
  pages={725},
  year={2025},
  publisher={Springer}
}

@article{porth2019event,
  title={The event horizon general relativistic magnetohydrodynamic code comparison project},
  author={Porth, Oliver and Chatterjee, Koushik and Narayan, Ramesh and Gammie, Charles F and Mizuno, Yosuke and Anninos, Peter and Baker, John G and Bugli, Matteo and Chan, Chi-kwan and Davelaar, Jordy and others},
  journal={The Astrophysical Journal Supplement Series},
  volume={243},
  number={2},
  pages={26},
  year={2019},
  publisher={IOP Publishing}
}

@article{daly2019black,
  title={Black hole spin and accretion disk magnetic field strength estimates for more than 750 active galactic nuclei and multiple galactic black holes},
  author={Daly, Ruth A},
  journal={The Astrophysical Journal},
  volume={886},
  number={1},
  pages={37},
  year={2019},
  publisher={IOP Publishing}
}

@article{eatough2013strong,
  title={A strong magnetic field around the supermassive black hole at the centre of the Galaxy},
  author={Eatough, RP and Falcke, H and Karuppusamy, R and Lee, KJ and Champion, DJ and Keane, EF and Desvignes, G and Schnitzeler, DHFM and Spitler, LG and Kramer, M and others},
  journal={Nature},
  volume={501},
  number={7467},
  pages={391--394},
  year={2013},
  publisher={Nature Publishing Group UK London}
}

@article{akiyama2021first,
  title={First M87 event horizon telescope results. VIII. Magnetic field structure near the event horizon},
  author={Akiyama, Kazunori and Algaba, Juan Carlos and Alberdi, Antxon and Alef, Walter and Anantua, Richard and Asada, Keiichi and Azulay, Rebecca and Baczko, Anne-Kathrin and Ball, David and Balokovi{\'c}, Mislav and others},
  journal={The Astrophysical Journal Letters},
  volume={910},
  number={1},
  pages={L13},
  year={2021},
  publisher={IOP Publishing}
}
\end{document}